\documentclass[onecollarge,natbib]{svjour2}
\bibpunct{[}{]}{;}{n}{}{,} 
\smartqed  
\usepackage{graphicx}
\usepackage{amsfonts}

\journalname{Few-Body Systems}
\begin{document}

\title{Three-body antikaon-nucleon systems
\thanks{The work was supported by the Czech CACR grant 15-04301S.}
}

\titlerunning{Three-body antikaon-nucleon systems}

\author{N.V. Shevchenko 
}


\institute{N.V. Shevchenko \at
              Nuclear Physics Institute, 25068 \v{R}e\v{z}, Czech Republic \\
              \email{shevchenko@ujf.cas.cz} 
}

\date{Received: date / Accepted: date}

\maketitle

\begin{abstract}
The paper contains a review of the exact or accurate results achieved in the field
of the three-body antikaon-nucleon physics. Different states and processes in
$\bar{K}NN$ and $\bar{K}\bar{K}N$ systems are considered. In particular,
quasi-bound states in $K^- pp$ and $K^- K^- p$ systems were investigated
together with antikaonic deuterium atom. Near-threshold scattering of antikaons
on deuteron, including the $K^- d$ scattering length, and applications of the scattering
amplitudes are also discussed.

All exact three-body results were calculated using some form of Faddeev equations.
Different versions of $\bar{K}N$, $\Sigma N$, $\bar{K}\bar{K}$, and $NN$ potentials,
specially constructed for the calculations, allowed investigation of the dependence of
the three-body results on the two-body input. Special attention is paid to the
antikaon-nucleon interaction, being the most important for the three-body systems.
Approximate calculations, performed additionally to the exact ones, demonstrate accuracy
of the commonly used approaches.

\end{abstract}

\tableofcontents

\section{Introduction}
\label{intro_sect}

An interest to the exotic systems, which consist of antikaons and nucleons, rose recently
after the statement \cite{AY1,AY2}, that deep and narrow quasi-bound states can exist
in $\bar{K}NN$ and $\bar{K}NNN$ systems. Due to this, several calculations of
the quasi-bound state in the lightest $\bar{K}NN$ system with $J^P = 0^-$ quantum numbers,
that is $K^- pp$, were performed. Among all calculations those using Faddeev-type equations
are the most accurate ones. The first results of the two accurate calculations
\cite{myKpp_PRL,myKpp_PRC,ikedasato_first} confirmed the existence of
the quasi-bound state in the $K^- pp$, but the evaluated binding energies and widths
are far different from those predicted in \cite{AY1,AY2}. The results of the two groups
\cite{myKpp_PRL,myKpp_PRC,ikedasato_first} also differ one from the other, and the main reason
for this is a choice of the antikaon-nucleon interaction, being an input for the three-body
calculations.

The question of the quasi-bound state in the $K^- pp$ system is far from being settled from
experimental point of view as well. The first experimental evidence of the $K^- pp$ quasi-bound state
existence occurred in the FINUDA experiment~\cite{FINUDA} at the DA$\Phi$NE $e^+ e^−$ collider.
Recently performed new analyses of old experiments, such as OBELIX~\cite{OBELIX}
at CERN and DISTO~\cite{DISTO} at SATURNE also claimed the observation of the state.
However, there are some doubts, whether the observed structure corresponds to the 
quasi-bound state. The experimental results also differ from each other, moreover, their
binding energies and widths are far from all theoretical predictions. Since the question
of the possible existence of the quasi-bound state in the $K^- pp$ system is still highly
uncertain, new experiments are being planned and performed by HADES~\cite{HADES} and LEPS~\cite{LEPS}
Collaborations, and in J-PARC E15~\cite{J-PARC_E15} and E27~\cite{J-PARC_E27} experiments.

It was demonstrated \cite{myKpp_PRC} that the $\bar{K}N$ interaction plays a crucial role
in the $\bar{K}NN$ calculations. It is much more important than the nucleon-nucleon
one, but is far less known. Therefore, a model of the antikaon-nucleon interaction,
which is more accurate in reproducing experimental data than that from \cite{myKpp_PRL,myKpp_PRC},
was necessary to construct. The experimental data on $\bar{K}N$ interaction,
which can be used for fitting parameters of the potential, are: near-threshold cross-sections
of $K^- p$ scattering, their threshold branching ratios, and shift and width
of $1s$ level of kaonic hydrogen (which should be more accurately called ``antikaonic
hydrogen''). The last observable has quite interesting experimental history and finally was
measured quite accurately. As for the theoretical description of kaonic hydrogen, many
authors used approximate Deser-type formulas, which connect $1s$ level shift
of an hadronic atom with the scattering length, given by strong interaction
in the pair. The question was, how accurate are the approximate formulas, derived for
the pion-nucleon interaction, for the antikaon-nucleon system. It was demonstrated in
\cite{revai_deser,our_KN} and later in \cite{cieply} that Deser-type formulas has low accuracy
for the antikaon-nucleon system.

Another question of antikaon-nucleon interaction is a structure of the $\Lambda(1405)$
resonance, which couples the $\bar{K}N$ system to the lower $\pi \Sigma$ channel.
$\Lambda(1405)$ is usually assumed as a quasi-bound state in the higher $\bar{K}N$ channel
and a resonance state in the lower $\pi \Sigma$ channel. But it was found in \cite{2L1405}
and later in other papers that a chirally motivated model of the interaction lead to
a two-pole structure of the resonance. Keeping these two points of view in mind,
two phenomenological $\bar{K}N$ potentials with one- and two-pole structure of the $\Lambda(1405)$
resonance were constructed in \cite{our_KN}. Their parameters were fitted to the experimental data,
and $1s$ level shift and width of kaonic hydrogen were calculated directly, without any
approximate formulas. 

It turned out \cite{our_KN} that it is possible to construct phenomenological potentials with
one- and two-pole $\Lambda(1405)$ resonance which describe existing low-energy
experimental data with the same level of accuracy. Due to this, the two $\bar{K}N$ potentials
were used as an input in calculations of the low-energy elastic $K^- d$ scattering in \cite{my_Kd}.
But after the publication of the $K^- d$ results, SIDDHARTA collaboration
reported results of their measurement of kaonic hydrogen characteristics \cite{SIDDHARTA}.
The results turned out to be quite different from the previously measured results of
DEAR experiment \cite{DEAR} and compatible with older KEK data \cite{KEK}. Due to this,
the $\bar{K}N$ potentials were refitted in such a way, that they reproduce the most recent
experimental data on the $1s$ level shift and width of kaonic hydrogen. The calculations of
the low-energy elastic kaon-deuteron scattering were repeated in \cite{my_Kd_sdvig} with
the new potentials. In addition, an approximate calculation of the $1s$ level shift and width
of kaonic deuterium was performed. It was done approximately using a complex $K^- - d$ potential,
reproducing the elastic three-body $K^- d$ amplitudes.

It was found that the three-body $K^- d$ system also does not allow to make preference to one
of the two phenomenological $\bar{K}N$ potentials and by this to solve the question
of the number of $\Lambda(1405)$ poles. In order to support the statement, one more,
a chirally motivated $\bar{K}N$ potential was constructed. As other chiral models, it has two poles
forming the $\Lambda(1405)$ resonance. Parameters of this potential were also fitted
to the low-energy experimental data on $K^- p$ scattering and kaonic hydrogen,
the chirally motivated $\bar{K}N$ potential reproduces all antikaon-nucleon data with
the same accuracy as the two phenomenological models.

Another way of investigation of the $\Lambda(1405)$ resonance was suggested and
realised in \cite{revai_1405}, were low-energy breakup of the $K^- d$ system was considered.
The idea was that the resonance should be seen as a bump in so called deviation spectrum of
neutrons in the final state of the reaction. However, $\Lambda(1405)$ is so broad that it
was seen as a bump in some cases only.

Finally, the calculations of the three-body $\bar{K}NN$ system with different quantum
numbers were repeated in \cite{ourKNN_I,ourKNN_II} using all three models of the $\bar{K}N$ interaction.
In particular, the binding energy and width of the $K^- pp$ quasi-bound state were evaluated,
the low-energy $K^- d$ amplitudes were calculated and the $1s$ level shift and width of kaonic
deuterium were predicted. A search of the quasi-bound state in the $K^- d$ system was also performed,
but the results are negative.

After the approximate calculations of the characteristics of deuterium, the exact
calculations were performed in \cite{our_Kdexact}. Namely, Faddeev-type equations with strong
plus Coulomb interactions, suggested in \cite{Papp1}, were solved.
It was the first time, when the equations \cite{Papp1}, initially written and used for a system
with Coulomb interaction being a correction to a strong potential, were used
for investigations of an hadronic atom, where Coulomb potential plays the main role.
Since the equations are much more complicated than ``usual'' AGS ones (containing short-range
potentials only), the calculations were performed with simple complex $\bar{K}N$ potentials,
reproducing only some of the experimental $K^- p$ data. Comparison of the dynamically exact
three-body results with the previous approximate ones shown that the approximation of the kaonic
deuterium as a two-body system is quite accurate for this task. 

Another three-body exotic system, consisting of two antikaons and one nucleon, was studied
in \cite{my_KKN}. It was expected that a quasi-bound state can exist in the $\bar{K}\bar{K}N$ system too.
The three $\bar{K}N$ potentials were used, and a quasi-bound state was found with smaller binding
energy than in the $K^- pp$ and larger width. It is interesting, that the parameters
of the state allow to associate it with a $\Xi$ state mentioned in the Particle Data Group \cite{PDG}.

The paper summarises results of the series of exact or accurate calculations
\cite{myKpp_PRL,myKpp_PRC,our_KN,my_Kd,my_Kd_sdvig,ourKNN_I,ourKNN_II,our_Kdexact,my_KKN}.
The next section contains information about the two-body interactions, necessary for the three-body
calculations. Faddeev-type Alt-Grassberger-Sandhas equations with coupled channels, which were used
for three-body calculations with strong interactions, are described in Section \ref{AGS_sect}.
Section \ref{qbs_sect} is devoted to the quasi-bound states in the $K^- pp$, $K^- d$, and $K^- K^- p$
systems. The near-threshold $K^-d$ scattering is considered in Section \ref{Kd_elastic_sect},
the kaonic deuterium -- in Section \ref{kaonic_deu_sect}. The last section summarises the results.

\section{Two-body interactions}
\label{Interactions_sect}

In order to investigate some three-body system it is necessary to know
the interactions of all the pairs of the particles. The interactions,
necessary for investigations of the $\bar{K}NN- \pi \Sigma N$ and
$\bar{K}\bar{K}N - \bar{K} \pi \Sigma$
systems, are $\bar{K}N$ and $\Sigma N$ with other channels coupled to them,
and the one-channel $NN$ and $\bar{K}\bar{K}$ interactions (the rest of them were
omitted in the three-body calculations). All potentials, except one of the $NN$ potentials, 
were specially constructed for the calculations. They have a separable form and
$\mathbb N$-term structure
\begin{equation}
\label{V_Nterm}
 V_{i,II'}^{\alpha \beta} = \sum_{m=1}^{{\mathbb N}_i^{\alpha}}
 \lambda_{i(m),II'}^{\alpha \beta} \,
 |g_{i(m),I}^{\alpha} \rangle  \langle g_{i(m),I'}^{\beta} |,
\end{equation}
which leads to a separable $T$-matrix
\begin{equation}
\label{T_Nterm}
 T_{i,II'}^{\alpha \beta} = \sum_{m,n = 1}^{{\mathbb N}_i^{\alpha}}
 |g_{i(m),I}^{\alpha} \rangle
\tau_{i(mn),II'}^{\alpha \beta} \langle g_{i(n),I'}^{\beta} | \,.
\end{equation}
${\mathbb N}_i^{\alpha}$ in Eq.~(\ref{V_Nterm}) and Eq.~(\ref{T_Nterm}) is a number of terms
of the separable potential, $\lambda$ is a strength constant, while $g$ is a form-factor.
The two-body isospin $I$ in general is not conserved. In particular, the two-body
isospin is conserved in the phenomenological $\bar{K}N-\pi \Sigma$ potentials, but
not in the corresponding $T$-matrices due to Coulomb interaction and the physical masses,
taken into account. The chirally motivated $\bar{K}N - \pi \Sigma - \pi \Lambda$ potential
does not conserve the isospin due to its energy and mass dependence.

The separable potentials are simpler than other models of interactions. However,
the potentials, entering the equations for the antikaon-nucleon systems, were constructed
in such a way that they reproduce the low-energy experimental data for every subsystem
very accurately. From this point of view they are not worse than other models of
the antikaon-nucleon or the $\Sigma$-nucleon interaction (in fact, they are even better
than some chiral models). The one-term NN potential does not have a repulsive part,
but the two-term model is repulsive at short distances. Finally, all three-body observables
described in the present paper turned out to be dependent on the $NN$ and $\Sigma N$
interactions very weakly, therefore the most important is the accuracy of the $\bar{K}N$
potential.

The antikaon-nucleon interaction is the most important one for the three-body systems
under consideration. There are several models of the $\bar{K}N$ interaction, some of them
are ``stand-alone'' ones having the only aim to reproduce experimental data, others were used
in few- of many-body calculations. The problem is that the first ones are too
complicated to be used in few-body calculations, while the models from the second group are
too simple to reproduce all the experimental data properly. Due to this, the $\bar{K}N$ potentials,
which are simple enough for using in Faddeev-type three-body equations and at the
same time reproduce all low-energy antikaon-nucleon experimental data, were constructed.

\subsection{Antikaon-nucleon interaction, experimental data}
\label{KN_exp_sect}

\underline{$\Lambda(1405)$ resonance}

The $\Lambda(1405)$ resonance is a manifestation of the attractive nature of the antikaon-nucleon
interaction in isospin zero state, it couples $\bar{K}N$ to the lower $\pi \Sigma$ channel.
Not only position and width, but the nature of the resonance itself are opened questions.
A usual assumption is that $\Lambda(1405)$ is a resonance in the $\pi \Sigma$ channel and
a quasi-bound state in the $\bar{K}N$ channel. According to the most recent Particle
Data Group issue~\cite{PDG}, the resonance has mass $1405.1^{+ 1.3}_{- 1.0}$ MeV
and width $50.5 \pm 2.0$ MeV. There is also an assumption suggested
in \cite{2L1405} and supported by other chiral models, that the bump, which
is usually understood as the $\Lambda(1405)$ resonance, is an effect of
two poles. Due to this, the two different phenomenological models of the antikaon-nucleon
interaction with one- or two-pole structure of the $\Lambda(1405)$ were constructed.
The third model is a chirally motivated potential, which has two poles by construction.

Extraction of the resonance parameters from experimental data is complicated for
two reasons. First, it cannot be studied in a two-body reaction and can be seen in
a final state of some few- or many-body process. Second, its width is large,
so the corresponding peak could be blurred.

A theoretical paper~\cite{revai_1405} was devoted to the possibility of tracing
the $\Lambda(1405)$ resonance in the neutron spectrum of a $K^- d$ breakup reaction.
The neutron spectra of the $K^- d \to \pi \Sigma n$ reaction were
calculated in center of mass energy range $0 - 50$ MeV. The three-body system with coupled
$\bar{K}NN$ and $\pi \Sigma N$ channels was studied using the Faddeev-type AGS equations, 
described in Section~{\ref{AGS_sect}}, with four phenomenological $\bar{K}N$
potentials with one- or two-pole structure of $\Lambda(1405)$.
It was found that kinematic effects completely mask the peak corresponding
to the $\Lambda(1405)$ resonance. Therefore, comparison of eventual experimental
data on the low-energy $K^- d \to \pi \Sigma n$ reaction with theoretical results
hardly can give an answer to the question of the number of $\Lambda(1405)$ poles. 

Later, similar calculations of the same process were performed for initial kaon
momentum $1$~GeV in \cite{Kdbreak_repeat1,Kdbreak_repeat2}. Coupled-channel AGS equations
were solved as well with energy-dependent and -independent $\bar{K}N$ potentials.
The authors predict a pronounced maximum in the double-differential cross section
with a forward emitted neutron at $\pi \Sigma$ invariant mass $1.45$ GeV. However,
applicability of the $\bar{K}N$ potentials, fitted to the near-threshold data,
and of the nonrelativistic Faddeev equations for such high energies is quite doubtful.

Several arguments, suggested in support to the idea of the two-pole structure
of the $\Lambda(1405)$ resonance, were checked in~\cite{my_Kd} using the one- and two-pole
phenomenological models of the antikaon-nucleon interaction. One of the arguments is
the difference between the $\pi \Sigma$ cross-sections with different charge combinations,
which is seen in experiments, e.g. in CLAS~\cite{CLAS}. The elastic $\pi^+ \Sigma^-$, $\pi^- \Sigma^+$,
and $\pi^0 \Sigma^0$ cross-sections were plotted to check the assumption, that
the difference is caused by the two-pole structure. However, it turned out that
the cross sections are different and their maxima are shifted one from each other
for both one- and two-pole versions of the $\bar{K}N$ potential (see Fig. 5 of ~\cite{my_Kd}).
Therefore, the effect is not a proof of the two-pole structure, but a manifestation of
the isospin non-conservation and differences in the background.

Another argument for the two-pole structure comes from the fact, that the poles in
a two-pole model are coupled to different channels. Indeed, a gradual switching
off of the coupling between the $\bar{K}N$ and $\pi \Sigma$ channels turns  
the upper pole into a real bound state in $\bar{K}N$, while the lower one becomes
a resonance in the uncoupled $\pi \Sigma$ channel (see e.g. Fig.2 of \cite{ourKNN_I}).
Consequently, it was suggested, that the poles of a two-body model
manifest  themselves in different reactions. In particular, the $\bar{K}N - \bar{K}N$,
$\bar{K}N - \pi \Sigma$, and $\pi \Sigma - \pi \Sigma$ amplitudes should
``feel'' only one of the two poles. The hypothesis was also checked in~\cite{my_Kd},
and indeed, the real parts of the $\bar{K}N - \bar{K}N$, $\bar{K}N - \pi \Sigma$, and
$\pi \Sigma - \pi \Sigma$ amplitudes in $I=0$ state cross the real axis at different energies.
But it is true for the both: the one- and the two-pole versions of the potential
(see Fig. 6 of \cite{my_Kd}). This effect must be caused by different background
contributions in the reactions. Therefore, a proof of the two-pole structure of
the $\bar{K}N$ interaction does not exist.

\vspace*{2mm}
\noindent
\underline{Cross-sections and threshold branching ratios}

Three threshold branching ratios of the $K^- p$ scattering 
\begin{eqnarray}
\label{gammaKp}
\gamma &=& \frac{\Gamma(K^- p \to \pi^+ \Sigma^-)}{\Gamma(K^- p \to
\pi^- \Sigma^+)} = 2.36 \pm 0.04, \\
\label{Rc}
R_c &=& \frac{\Gamma(K^- p \to \pi^+ \Sigma^-, \pi^- \Sigma^+)}{\Gamma(K^- p \to
\mbox{all inelastic channels} )} = 0.664 \pm 0.011, \\
\label{Rn}
R_n &=& \frac{\Gamma(K^- p \to \pi^0 \Lambda)}{\Gamma(K^- p \to
\mbox{neutral states} )} = 0.189 \pm 0.015
\end{eqnarray}
were measured rather accurately in~\cite{gammaKp1, gammaKp2}. Since the phenomenological
$\bar{K}N$ potentials, used in the calculations, take the lowest
$\pi \Lambda$ channel into account indirectly, a new ratio
\begin{equation}
\label{RpiSigma}
 R_{\pi \Sigma} =
 \frac{\Gamma(K^- p \to \pi^+ \Sigma^-)+\Gamma(K^- p \to \pi^- \Sigma^+)}{
 \Gamma(K^- p \to \pi^+ \Sigma^-) + \Gamma(K^- p \to \pi^- \Sigma^+) +
                                   \Gamma(K^- p \to \pi^0 \Sigma^0) } \,,
\end{equation}
which contains the measured $R_c$ and $R_n$ and has an ``experimental'' value
\begin{equation}
\label{RpiSigma_exp}
 R_{\pi \Sigma} =  \frac{R_c}{1-R_n \, (1 - R_c)} \, = \, 0.709 \pm 0.011
\end{equation}
was constructed and used.

In contrast to the branching ratios, the elastic and inelastic total cross sections
with $K^- p$ in the initial state~\cite{Kp2exp,Kp3exp_1,Kp3exp_2,Kp4exp,Kp5exp,Kp6exp}
were measured not so accurately, see Figure~\ref{CrossKp.fig}.

\vspace*{2mm}
\noindent
\underline{Kaonic hydrogen}

The most promising source of knowledge about the $\bar{K}N$ interaction is kaonic hydrogen
atom (which correctly should be called ''antikaonic hydrogen''). The atom has rich
experimental history, several experiments measured its $1s$ level shift
\begin{equation}
\Delta E_{1s} = E_{1s}^{Coul} - {\rm Re }(E_{1s}^{Coul+Strong})
\end{equation}
and width $\Gamma_{1s}$, caused by the strong $\bar{K}N$ interaction in comparison to pure
Coulomb case, with quite different results. The most recent measurement was performed by
SIDDHARTA collaboration~\cite{SIDDHARTA}, their results are:
\begin{equation}
\label{SIDDHARTA}
\Delta E_{1s}^{\rm SIDD} = -283 \pm 36 \pm 6 \; {\rm eV}, \quad
\Gamma_{1s}^{\rm SIDD} = 541 \pm 89 \pm 22 \; {\rm eV}.
\end{equation}
Paradoxically, the directly measurable observables are not reproduced in the same
way in the most of the theoretical works devoted to the antikaon-nucleon interaction.
Some approximate formula are usually used for reproducing the $1s$ level shift.
The most popular is a ``corrected Deser'' formula~\cite{corDeser}, which connects
the shift with the scattering length $a_{K^- p}$ of the $K^- p$ system:
\begin{eqnarray}
\label{corDes}
  \Delta E^{cD} - i \, \frac{\Gamma^{cD}}{2} 
      &=& - 2 \alpha^3 \mu_{K^-d}^2 \, a_{K^-p} \\
  \nonumber
  &{}& \times [1 - 2 \alpha \mu_{K^-p} \, a_{K^- p} \, (\ln \alpha - 1)].
\end{eqnarray}
The formula is one of quite a few versions of the original formula, derived
by Deser for the pion-nucleon system. It differs from the original one by the second term
in the brackets. However, it was shown (e.g. in \cite{our_KN} and other papers) that
for the antikaon-nucleon system the formula is not accurate, it gives $\sim 10\%$ error.

\subsection{Phenomenological and chirally motivated $\bar{K}N$ potentials}
\label{V_KN_sect}

The constructed phenomenological models of antikaon-nucleon interaction with one- or
two-pole structure of the $\Lambda(1405)$ resonance together with the chirally motivated
model reproduce the $1s$ level shift and width of kaonic hydrogen, measured
by SIDDHARTA collaboration, directly, without using approximate formulas. The potentials
also reproduce the experimental data on the $K^- p$ scattering and the threshold branching
ratios, described in the previous subsection. All three potentials are suitable for using
in accurate few-body equations.

The problem of two particles interacting by the strong and Coulomb potentials,
considered on the equal basis, was solved. The method of solution of Lippmann-Schwinger
equation for a system with Coulomb plus a separable strong potential is based on the fact
that the full $T$-matrix of the problem can be written as a sum $T = T^c + T^{sc}$.
Here $T_c$ is the pure Coulomb transition matrix and $T^{sc}$ is the Coulomb-modified
strong $T$-matrix. It was necessary to extend the formalism to describe the system
of the coupled $\bar{K}N$,  $\pi \Sigma$ (and $\pi \Lambda$ for the chirally
motivated potential) channels. The physical masses of the particles were used in the equation,
therefore, the two-body isospin of the system is not conserved. More details on the formalism
can be found in~\cite{our_KN}.
\begin{center}
\begin{table}
\caption{Physical characteristics of the three antikaon nucleon potentials:
phenomenological $V^{1,SIDD}_{\bar{K}N-\pi \Sigma}$ and
$V^{2,SIDD}_{\bar{K}N-\pi \Sigma}$ with one- and two-pole structure of
the $\Lambda(1405)$ resonance respectively, and the chirally
motivated $V^{Chiral}_{\bar{K}N - \pi \Sigma-\pi \Lambda}$ potential: 
$1s$ level shift $\Delta E_{1s}^{K^- p}$ (eV) and
width $\Gamma_{1s}^{K^- p}$ (eV) of kaonic hydrogen,
threshold branching ratios $\gamma$, $R_c$ and $R_n$ together
with the experimental data. The additional $R_{\pi \Sigma}$ ratio, see Eq.(\ref{RpiSigma}),
with its ``experimental'' value is shown as well. Scattering length of the $K^- p$
system $a_{K^- p}$ (fm) and pole(s) $z_1$, $z_2$ (MeV) forming the $\Lambda(1405)$ resonance
are also demonstrated.}
\label{phys_char.tab}
\begin{tabular}{ccccc}
\hline \noalign{\smallskip}
 & $V^{1,SIDD}_{\bar{K}N-\pi \Sigma}$ & $V^{2,SIDD}_{\bar{K}N-\pi \Sigma}$ &
  $V^{Chiral}_{\bar{K}N - \pi \Sigma-\pi \Lambda}$ & Experiment  \\
\noalign{\smallskip} \hline \noalign{\smallskip}
$\Delta E_{1s}^{K^- p}$  & -313 & -308 & $-313$  & $-283 \pm 36 \pm 6$~\cite{SIDDHARTA} \\
$\Gamma_{1s}^{K^- p}$    & 597 & 602 & $561$ & $541 \pm 89 \pm 22$~\cite{SIDDHARTA} \\
$\gamma$ & 2.36 & 2.36 & 2.35 & $2.36 \pm 0.04$~\cite{gammaKp1,gammaKp2} \\
$R_c$ & - & - & 0.663 & $0.664 \pm 0.011$~\cite{gammaKp1,gammaKp2} \\
$R_n$ & - & - & 0.191 & $0.189 \pm 0.015$~\cite{gammaKp1,gammaKp2} \\
\noalign{\smallskip} \hline \noalign{\smallskip}
$R_{\pi \Sigma}$ & 0.709 & 0.709 & - & $0.709 \pm 0.011$~Eq.(\ref{RpiSigma_exp}) \\
$a_{K^- p}$ & -0.76 + i 0.89 & -0.74 + i 0.90 & -0.77 + i 0.84 &  \\
$z_1$ & 1426 - i 48 & 1414 - i 58  & 1417 - i 33 &  \\
$z_2$ &  -          & 1386 - i 104 & 1406 - i 89 &  \\
\noalign{\smallskip} \hline
\end{tabular}
\end{table}
\end{center}

The phenomenological potentials describing the $\bar{K}N$ system with coupled
$\pi \Sigma$ channel are the one-term separable ones defined by~Eq.(\ref{V_Nterm}).
In momentum representation they have a form
\begin{equation}
\label{Vseprb}
 V_{I}^{\bar{\alpha} \bar{\beta}}(k^{\bar{\alpha}},k'^{\bar{\beta}}) =
 \lambda_{I}^{\bar{\alpha} \bar{\beta}} \;
 g^{\bar{\alpha}}(k^{\bar{\alpha}}) \, g^{\bar{\beta}}(k'^{\bar{\beta}}),
\end{equation}
where indices $\bar{\alpha}, \bar{\beta} = 1, 2$ denote the $\bar{K}N$
or $\pi \Sigma$ channel respectively, and $I$ is a two-body isospin.
Different form-factors were used for the one- and two-pole versions
of the phenomenological potential. While for the one-pole version
Yamaguchi form-factors
\begin{equation}
\label{1res_ff}
 g^{\bar{\alpha}}(k^{\bar{\alpha}}) = \frac{1}{(k^{\bar{\alpha}})^2 +
 (\beta^{\bar{\alpha}})^2}
\end{equation}
were used, the two-pole version has slightly more complicated form-factors
in the $\pi \Sigma$ channel
\begin{equation}
\label{2res_ffpi}
 g^{\bar{\alpha}}(k^{\bar{\alpha}}) = 
 \frac{1}{(k^{\bar{\alpha}})^2
   + (\beta^{\bar{\alpha}})^2} \,+\,
 \frac{s \, (\beta^{\bar{\alpha}})^2}{[(k^{\bar{\alpha}})^2 + 
 (\beta^{\bar{\alpha}})^2]^2} \,.
\end{equation}
In the $\bar{K}N$ channel the form-factor of the two-pole version is also
of Yamaguchi form Eq.(\ref{1res_ff}). 

Range parameters $\beta^{\bar{\alpha}}$,
strength parameters $\lambda_{I}^{\bar{\alpha} \bar{\beta}}$
and an additional parameter $s$ of the two-pole version were obtained by
fitting to the experimental data described in the previous subsection.
They are: the elastic and inelastic $K^- p$ cross-sections, the threshold branching
ratios and the $1s$ level shift and width of kaonic hydrogen. The first versions
of the potentials, presented in~\cite{our_KN} and used in \cite{my_Kd}, were
fitted to the KEK data \cite{KEK} on kaonic hydrogen. The actual versions of the
phenomenological potentials were fitted to the most recent experimental data of
SIDDHARTA collaboration \cite{SIDDHARTA}. The parameters of the one- and
two-pole versions of the phenomenological potentials fitted to SIDDHARTA data
can be found in~\cite{my_Kd_sdvig}.

All fits were performed directly to the experimental values except the threshold
branching ratios $R_c$ and $R_n$. The reason is that the ratios contain data on scattering
of $K^- p$ into all inelastic channels including $\pi \Lambda$, which is taken 
by the phenomenological potentials into account
only indirectly through imaginary part of one of the $\lambda$ parameters. Due
to this the phenomenological potentials were fitted to the new ratio $R_{\pi \Sigma}$,
defined in Eq.(\ref{RpiSigma_exp}).

The third model of the antikaon-nucleon interaction is the chirally motivated potential.
It connects all three open channels: $\bar{K}N$, $\pi \Sigma$ and $\pi \Lambda$,
and has a form
\begin{equation}
\label{VchiralIso}
V_{I I'}^{\alpha \beta}(k^{\alpha},k'^{\beta};\sqrt{s} ) =
  g_I^{\alpha} (k^{\alpha}) \, \bar{V}_{I I'}^{\alpha \beta}(\sqrt{s}) \,
  g_{I'}^{\beta} (k'^{\beta}) \,,
\end{equation}
where $V_{I I'}^{\alpha \beta}(\sqrt{s})$ is the energy dependent part of the
potential in isospin basis. In particle basis the energy dependent part
has a form
\begin{equation}
\label{VchiralPart}
\bar{V}^{ab}(\sqrt{s} ) =
  \sqrt{ \frac{M_{a}}{2 \omega_{a} E_{a} }} \,
  \frac{C^{ab}(\sqrt{s})}{(2 \pi)^3 f_{a} f_{b}} \,
  \sqrt{ \frac{M_{b}}{2 \omega_{b} E_{b} }} \,.
\end{equation}
Indices $a, b$ here denote the particle channels
$a, b = K^- p, \bar{K}^0n, \pi^+ \Sigma^-, \pi^0 \Sigma^0, \pi^- \Sigma^+$
and $\pi^0 \Lambda$.  the square roots with baryon mass $M_{a}$, baryon energy
$E_{a}$ and meson energy $w_{a}$ of the channel $a$ ensure proper normalization
of the corresponding amplitude. SU(3) Clebsh-Gordan coefficients $C^{WT}_I$
enter the non-relativistic form of the leading order Weinberg-Tomozawa interaction
\begin{equation}
C^{ab}(\sqrt{s}) = 
 - C^{WT} \, (2\sqrt{s} - M_{a} - M_{b}).
\end{equation}
Since, as in the case of the phenomenological potentials, the physical
masses of the particles were used, the two-body isospin $I$ in Eq.(\ref{VchiralIso})
is not conserved. It is different from the phenomenological potentials situation since
in that case the potentials conserve the two-body isospin (but the corresponding
$T$-matrices do not). Another feature, which distinguish the chirally motivated
potential from the phenomenological ones, is the isospin dependence of its
form-factors:
\begin{equation}
g_I^{\alpha} (k^{\alpha}) = 
 \frac{(\beta_I^{\alpha})^2}{(k^{\alpha})^2 + (\beta_I^{\alpha})^2}.
\end{equation}
Besides, they are dimensionless due to the additional factor $(\beta_I^{\alpha})^2$ in
the numerator.

The pseudo-scalar meson decay constants $f_{\pi}$, $f_K$ and the isospin dependent
range parameters $\beta_I^{\alpha}$ are free parameters of the chirally motivated potential.
They also were found by fitting the potential to the experimental data in the same way as
in the phenomenological potentials case. The chirally-motivated
$\bar{K}N - \pi \Sigma - \pi \Lambda$ potential reproduces the elastic and inelastic
$K^- p$ cross-sections, SIDDHARTA $1s$ level shift and width of kaonic hydrogen. In contrast
to the phenomenological potentials, the chirally-motivated one directly reproduces all three
$K^- p$ branching ratios: $\gamma$, $R_c$ and $R_n$. The parameters of the potential
can be found in~\cite{ourKNN_I}.

The $\Lambda(1405)$ resonance can manifest itself as a bump in elastic $\pi \Sigma$
cross-sections or in $K^- p$ amplitudes. In the last case, the real part of the
amplitude crosses zero, while the imaginary part has a maximum near the resonance
position. It is demonstrated in \cite{my_Kd_sdvig} and \cite{ourKNN_I} that
the elastic $\pi \Sigma$ cross-sections, provided by the three potentials, have a bump
near the PDG value \cite{PDG} for the mass of the $\Lambda(1405)$ resonance
with appropriate width.
\begin{figure}
\centering
\includegraphics[width=0.9\textwidth]{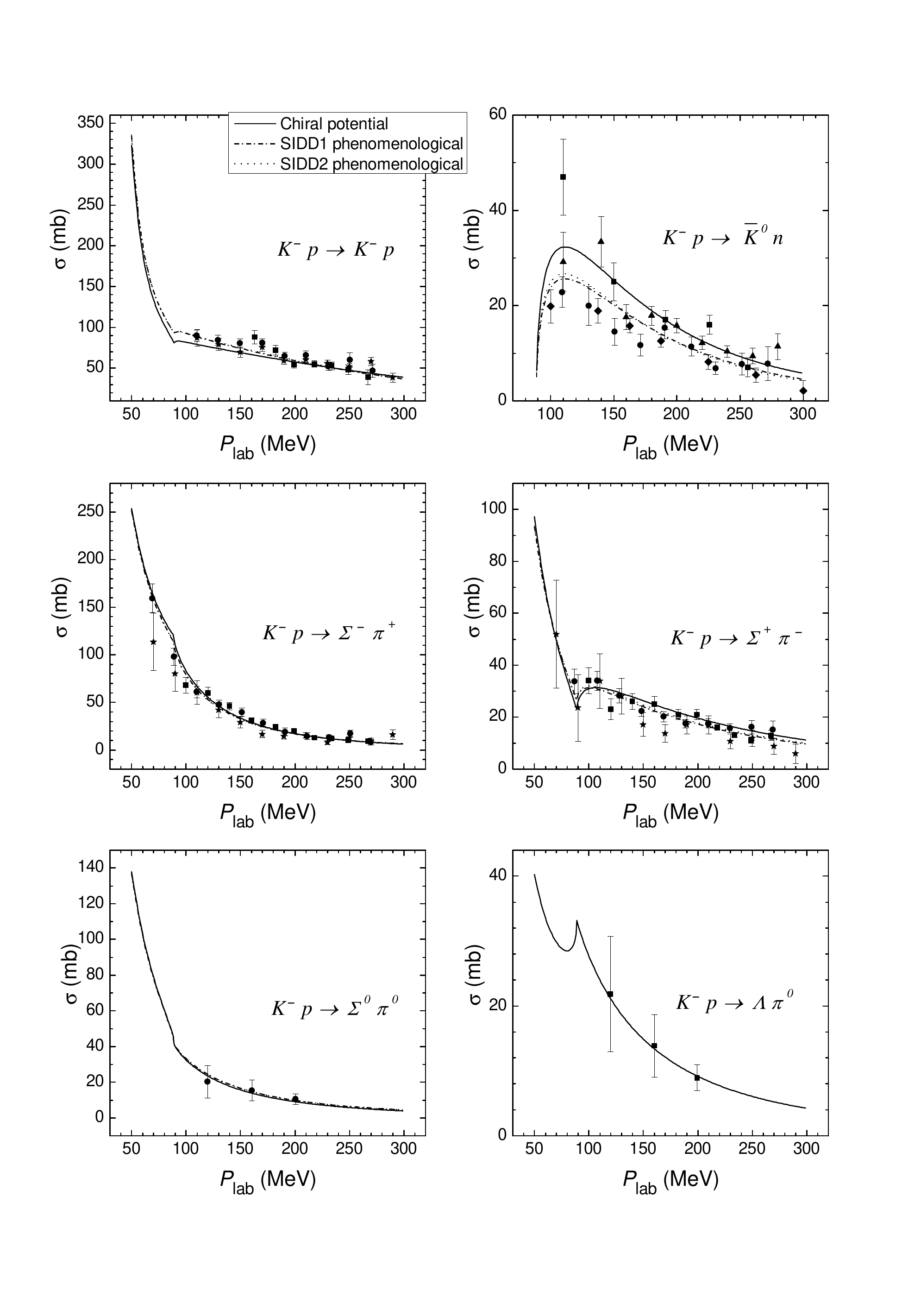}
\caption{Elastic and inelastic $K^- p$ cross-sections, obtained with
the one-pole $V^{\rm 1, SIDD}_{\bar{K}N - \pi \Sigma}$ (dash-dotted line),
the two-pole $V^{\rm 2, SIDD}_{\bar{K}N - \pi \Sigma}$ (dotted line)
phenomenological potentials,
and the chirally motivated $V^{Chiral}_{\bar{K}N - \pi \Sigma-\pi \Lambda}$
(solid line) potential. The experimental data are taken
from~\cite{Kp2exp,Kp3exp_1,Kp3exp_2,Kp4exp,Kp5exp,Kp6exp}.
\label{CrossKp.fig}}
\end{figure}

The physical characteristics of the three antikaon-nucleon potentials are shown in
Table~\ref{phys_char.tab}. In addition, the $K^- p$ scattering length $a_{K^- p}$
and the pole(s) forming the $\Lambda(1405)$ resonance, given by the potentials,
are demonstrated. The elastic and inelastic $K^- p$ cross-sections, provided by
the three potentials, are plotted in Figure \ref{CrossKp.fig} together with the
experimental data. It is seen form the Table~\ref{phys_char.tab} and
Figure \ref{CrossKp.fig}  that the one- and two-pole phenomenological potentials
and the chirally motivated potential describe the all experimental
data with equal high accuracy. Therefore, it is not possible to choose one of
the three models of the $\bar{K}N$ interaction looking at the two-body system.

\vspace*{2mm}
\noindent \underline{Approximate versions of the coupled-channel potential}

In order to check approximations used in other theoretical works, two approximate
versions of the coupled-channel potentials~Eq.(\ref{Vseprb}), which have only one
$\bar{K}N$ channel, were also used. They are: an exact optical potential and
a simple complex one.

The exact optical one-channel potential, corresponding to a two-channel one,
is given by Eq.~(\ref{Vseprb}) with $\bar{\alpha}, \bar{\beta} = 1$ and
the strength parameter defined as
\begin{equation}
\label{lambdaOpt}
\lambda^{11,{\rm Opt}}_{I} = \lambda^{11}_{I} +
\frac{(\lambda^{12}_{I})^2 \,
\langle g^{2}_{I} |\, G_0^{(2)}(z^{(2)}) |\, g^{2}_{I} \rangle}{
1 - \lambda^{22}_{I} \, \langle g^{2}_{I} |\, G_0^{(2)}(z^{(2)}) |\, g^{2}_{I} \rangle
} \,.
\end{equation}
Here $\lambda^{\bar{\alpha},\bar{\beta}}_I$ are the strength parameters of the two-channel
potential, and $| g^{2}_{I} \rangle$ is the form-factor of the second channel.
Since a two-body free Green's function $G_0^{(2)}$ depends on the corresponding
two-body energy, the parameter $\lambda^{11,{\rm Opt}}_{I}$ of the exact optical
potential is an energy-dependent complex function. The exact optical potential has
exactly the same elastic amplitudes of the $\bar{K}N$ scattering as the elastic part
of the full potential with coupled channels.

A simple complex potential is quite often miscalled ``an optical'' one, however, it is 
principally different. The strength parameter $\lambda^{11,{\rm Complex}}_{I}$
of a simple complex potential is a complex constant, therefore, the simple complex
potential is energy independent. The strength parameter of a simple complex potential
is usually chosen in such a way, that the potential reproduces only some characteristics
of the interaction. The simple complex potential as well as the exact optical one take
into account flux losses into inelastic channels through imaginary parts of the strength
parameters.

\subsection{Nucleon-nucleon and $\Sigma N (-\Lambda N)$ potentials}
\label{NN_SigN_sect}

\noindent \underline{$NN$ interaction}

Different $NN$ potentials, in particular, TSA-A, TSA-B and PEST, were used in order
to investigate dependence of the three-body results on the nucleon-nucleon interaction models.

A two-term separable $NN$ potential~\cite{Doles_NN}, called TSA, reproduces
Argonne $V18$~\cite{ArgonneV18} phase shifts and, therefore, is repulsive at
short distances. Two versions of the potential (TSA-A and TSA-B) with
slightly different form-factors
\begin{eqnarray}
&{}& g_{(m)}^{A,NN}(k) = \sum_{n=1}^2 \frac{\gamma_{(m)n}^A}{(\beta_{(m)n}^A)^2 + k^2},
\quad {\rm for \;} (m)=1,2 \\
\nonumber
&{}& g_{(1)}^{B,NN}(k) = \sum_{n=1}^3 \frac{\gamma_{(1)n}^B}{(\beta_{(1)n}^B)^2 + k^2},
\quad
g_{(2)}^{B,NN}(k) = \sum_{n=1}^2 \frac{\gamma_{(2)n}^B}{(\beta_{(2)n}^B)^2 + k^2}
\end{eqnarray}
were used. TSA-A and TSA-B potentials properly reproduce the $NN$ scattering lengths and
effective radii, they also give correct binding energy of the deuteron in the ${}^3S_1$ state.
For more details see Ref. \cite{my_Kd}.

A separabelization of the Paris model of the $NN$ interaction, called PEST potential~\cite{NNpot},
was also used. The strength parameter of the one-term PEST is equal to $-1$,
while the form-factor is defined by
\begin{equation}
g_{I}^{NN}(k) = \frac{1}{2 \sqrt{\pi}} \, \sum_{n=1}^6
\frac{c_{n,I}^{NN}}{k^2 + (\beta_{n,I}^{NN})^2}
\end{equation}
with $c_{n,I}^{NN}$ and $\beta_{n,I}^{NN}$ being the parameters.
PEST is equivalent to the Paris potential on and off energy shell
up to $E_{\,\rm lab} \sim 50$ MeV. It also reproduces the deuteron binding
energy in the ${}^3S_1$ state, as well as the triplet and singlet $NN$
scattering lengths.

The quality of reproducing the ${}^3S_1$ phase shifts by the three $NN$ potentials
is shown in Fig.8 of \cite{my_Kd}, were they are compared with those given by the Argonne V18
model. The two-term TSA-A and TSA-B potentials are very good at reproducing of
the Argonne V18 phase shifts. They cross the real axis, which is a consequence of the $NN$
repulsion at short distances. The one-term PEST potential does not have such a property,
but its phase shifts are also close to the ``standard'' ones at lower energies.

Only isospin-singlet $NN$ potential enters the AGS equations for the $K^- d$ system and
only isospin-triplet one enters the equations describing the $K^- pp$ system after
antisymmetrization.

\vspace*{2mm}
\noindent \underline{$\Sigma N$ interaction}

A spin dependent $V^{\rm Sdep}$ and an independent of spin $V^{\rm Sind}$ versions
of the $\Sigma N$ potential were constructed in \cite{my_Kd} in such a way, that
they reproduce the experimental $\Sigma N$ and $\Lambda N$
cross-sections~\cite{SigmaN1,SigmaN2,SigmaN3,SigmaN4,SigmaN5}.
The one-term separable potentials with Yamaguchi form-factors were used for the
two possible isospin states, but with different number of the channels.

Parameters of the one-channel $I=\frac{3}{2}$ state were fitted to the
$\Sigma^+ p \to \Sigma^+ p$ cross-sections. The $\Sigma N$ system in isospin one half
state is connected to the $\Lambda N$ channel, therefore, a coupled-channel potential
of the $I=\frac{1}{2}$ $\Sigma N - \Lambda N$ interaction was constructed first.
The coupled-channel $I=\frac{1}{2}$ potential together with the one-channel $I=\frac{3}{2}$
potential reproduce the $\Sigma^- p \to \Sigma^- p$, $\Sigma^- p \to \Sigma^0 n$,
$\Sigma^- p \to \Lambda n$, and $\Lambda p \to \Lambda p$ cross-sections.
It is seen in Fig.9 of \cite{my_Kd} that both $V^{\rm Sdep}$ and $V^{\rm Sind}$
versions of the $I=3/2$ $\Sigma N$ and $I=1/2$ $\Sigma N - \Lambda N$ potentials
reproduce the experimental data perfectly. Parameters of the potentials and
the scattering lengths $a^{\Sigma N}_{\frac{1}{2}}$, $a^{\Sigma N}_{\frac{3}{2}}$,
and $a^{\Lambda N}_{\frac{1}{2}}$, given by them, are shown in Table 5 of \cite{my_Kd}.

For the three-body $\bar{K}NN$ calculations, where a channel containing $\Lambda$ is not
included directly, not a coupled-channel, but one-channel $\Sigma N$ models of
the interaction in the $I=\frac{1}{2}$ state were used. They are an exact
optical $V^{\Sigma N, \rm{Opt}}$ potential and a simple complex $V^{\Sigma N, \rm{Complex}}$
one, corresponding to the $I=\frac{1}{2}$ $\Sigma N - \Lambda N$ model
with coupled channels. The exact optical potential has an energy dependent strength
parameter defined by Eq.~(\ref{lambdaOpt}), it reproduces the elastic $\Sigma N$ amplitude
of the corresponding two-channel potential exactly. The simple complex potential gives
the same scattering lengths, as the two-channel potential.

\subsection{Antikaon-antikaon interaction}
\label{V_KK_sect}

Lack of an experimental information on the $\bar{K}\bar{K}$ interaction
means that it is not possible to construct the $\bar{K}\bar{K}$ potential in the same
way as the $\bar{K}N$ or $\Sigma N$ ones. Due to this, theoretical results 
of a modified model describing the $\pi\pi - K\bar{K}$ system developed by the J\"ulich
group \cite{Lohse90,Janssen95} were used. The original model yields a good description of
the $\pi\pi$ phase shifts up to partial waves with total angular momentum $J=2$ and for
energies up to $z_{\pi\pi}\approx 1.4$ GeV. In addition, the $f_0(980)$ and $a_0(980)$
mesons result as dynamically generated states.

Based on the underlying SU(3) flavor symmetry, the interaction in 
the $\bar{K}\bar{K}$ system was directly deduced from the $K\bar{K}$ interaction
without any further assumptions. The $\bar{K}\bar{K}$ scattering length
predicted by the modified J\"ulich model is $a_{\bar{K}\bar{K},I=1}= -0.186$ fm,
therefore, it is a repulsive interaction. This version of the $\bar{K}\bar{K}$
interaction was called ''Original''.

Recent results for the $KK$ scattering length from lattice QCD simulations
suggest values of $a_{\bar{K}\bar{K},I=1}= (-0.141\pm 0.006)$ fm \cite{Beane:2007uh}
and $a_{\bar{K}\bar{K},I=1}= (-0.124\pm 0.006\pm 0.013)$ fm \cite{Sasaki:2013vxa}.
Those absolute values are noticeably smaller than the one predicted by the Original
J\"ulich meson-exchange model and, accordingly, imply a somewhat less repulsive
$\bar{K}\bar{K}$ interaction. Due to this, another version of the interaction that is
in line with the lattice QCD results was also constructed. It yields scattering length
$a_{\bar{K}\bar{K},I=1}= -0.142$ fm. This version of the $\bar{K}\bar{K}$ interaction
was called ''Lattice motivated''.

However, the models of the $\bar{K}\bar{K}$ interaction  described above cannot be
directly used in the AGS equations. Due to this, the $\bar{K}\bar{K}$ interaction
was represented in a form of the one-term separable potential with form factors given by 
\begin{equation}
 g(k) = \frac{1}{\beta_1^2 + k^2} +
  \frac{\gamma}{\beta_2^2 + k^2}.
\end{equation}
The strength parameters $\lambda$, $\gamma$ and range parameters
$\beta$ were fixed by fits to the $\bar{K}\bar{K}$ phase shifts and
scattering lengths of the ''Original'' and the `'Lattice motivated'' 
models of the antikaon-antikaon interaction.

\section{AGS equations for coupled $\bar{K}NN - \pi \Sigma N$ and
$\bar{K}\bar{K}N - \bar{K}\pi\Sigma$ channels}
\label{AGS_sect}

The three-body Faddeev equations in the Alt-Grassberger-Sandhas (AGS)
form~\cite{AGS} were used for the most of the three-body calculations.
The equations were extended in order to take the $\pi \Sigma$ channel, coupled to
the $\bar{K}N$ subsystem, directly. In practice it means that all operators
entering the system
\begin{equation}
\label{U_coupled}
U_{ij}^{\alpha \beta} = {\delta}_{\alpha \beta} \,(1-\delta_{ij})
\, \left(G_0^{\alpha} \right)^{-1} + \sum_{k, \gamma=1}^3 (1-\delta_{ik})
\, T_k^{\alpha \gamma} \, G_0^{\gamma} \, U_{kj}^{\gamma \beta} \, ,
\end{equation}
namely, transition operators $U_{ij}$, two-body $T$-matrices $T_i$
and the free Green function $G_0$, - have additional channel indices $\alpha, \beta = 1,2,3$
in addition to the Faddeev partition indices $i,j = 1,2,3$. 
The additional $\pi \Sigma N$ ($\alpha=2$) and $\pi N \Sigma$ ($\alpha=3$) channels
were added to the $\bar{K}NN$ system , while the $\bar{K}\bar{K}N$ system
was extended to the $\bar{K} \pi \Sigma$ ($\alpha=2$) and $\pi \bar{K} \Sigma$
($\alpha=3$) channels. A Faddeev index $i$, as usual, defines a particle and
the remained pair, now in the particular particle channel $\alpha$. The combinations
of the $(i,\alpha)$ indices with possible two-body isospin values can be found
in~\cite{myKpp_PRC} for the $\bar{K}NN$ system and in~\cite{my_KKN} for
the $\bar{K}\bar{K}N$ systems, respectively.

Since the separable potentials Eq.(\ref{V_Nterm}), leading to the separable $T$-matrices
Eq.(\ref{T_Nterm}), were used as an input, the system Eq.(\ref{U_coupled})
turned into the new system of operator equations
\begin{equation}
\label{full_oper_eq}
X_{ij, I_i I_j}^{\alpha \beta} = \delta_{\alpha \beta} \,
Z_{ij, I_i I_j}^{\alpha} +
\sum_{k=1}^3 \sum_{\gamma=1}^3 \sum_{I_k}
Z_{ik, I_i I_k}^{\alpha} \, \tau_{k, I_k}^{\alpha \gamma} \,
X_{kj, I_k I_j}^{\gamma \beta}
\end{equation}
with $X_{ij, I_i I_j}^{\alpha \beta}$ and $Z_{ij, I_i I_j}^{\alpha \beta}$
being new transition and kernel operators respectively
\begin{eqnarray}
\label{X_definition}
X_{ij, I_i I_j}^{\alpha \beta} &=&  \langle
g_{i,I_i}^{\alpha} | G_0^{\alpha} \, U_{ij, I_i I_j}^{\alpha
\beta} G_0^{\beta} | g_{j,I_j}^{\beta} \rangle \,, \\
\label{Z_definition}
Z_{ij, I_i I_j}^{\alpha \beta} &=&
\delta_{\alpha \beta} \, Z_{ij, I_i I_j}^{\alpha} =
\delta_{\alpha \beta} \, (1-\delta_{ij}) \,
\langle g_{i,I_i}^{\alpha} | G_0^{\alpha} | g_{j,I_j}^{\alpha}
\rangle \, .
\end{eqnarray}

\vspace*{2mm}
\noindent
\underline{$\bar{K}NN - \pi \Sigma N$ system}

The two states of the strangeness $S=-1$ $\bar{K}NN$ system were considered.
The $K^- pp$ and $K^- d$ systems differ from each another by the total spin value,
which leads to different symmetry of the operators describing the system containing
two identical baryons, $NN$. This fact is taken into account when the three-body
coupled-channel equations are antisymmetrized.

All calculations were performed under or slightly above the $\bar{K}NN$ threshold,
so that orbital angular momentum of all two-body interactions was set to zero
and, therefore, the total orbital angular momentum is also $L=0$. In particular,
the main $\bar{K}N$ potential was constructed with orbital angular momentum $l=0$ since
the interaction is dominated by the $s$-wave $\Lambda(1405)$ resonance. The interaction
of $\pi$-meson with a nucleon is weaker than
the other interactions, therefore, it was omitted in the equations.
An experimental information about the $\Sigma N$ interaction is very poor, and there
is no reason to assume significant effect of higher partial waves. Finally, the $NN$ interaction 
was also taken in $l=0$ state only, since physical reasons for
sufficient effect of higher partial waves in the present calculation are not seen.

Spin of the $\bar{K}NN$ system is given by spin of the two baryons, which also defines
the $NN$ isospin due to the symmetry properties. Looking for the quasi-bound state
in $\bar{K}NN$, the isospin $I=1/2$ and spin zero state, usually denoted as $K^- pp$,
was chosen due to its connection to experiment. Another possible configuration
with the same isospin and spin one, which is $K^- d$, was also studied. As for
the $\bar{K}NN$ state with isospin $I=3/2$, it is governed by the isospin $I_i=1$
$\bar{K}N$ interaction, which is much weaker attractive than the one in the $I_i=0$ state
or even repulsive. Therefore, no quasi-bound state is expected there.

The nucleons, entering the highest $\bar{K}NN$ channel, require
antisymmetrization of the operators entering the system of equations~Eq.(\ref{full_oper_eq}). 
Two identical baryons with symmetric spatial components ($L_i=0$) has
antisymmetric ($S_i=0$) spin components for the $pp$ state of the $NN$ subsystem
or symmetric ($S_i=1$) ones for the $d$ state. 
The operator $X_{1,1}^1$ has symmetric $NN$ isospin components,
therefore, it has the correct symmetry properties for the $K^- pp$
system (here and in what follows the right-hand indices of
$X$ are omitted: $X_{ij,I_i I_j}^{\alpha \beta} \to X_{i,I_i}^{\alpha}$).
Another operator, $X_{1,0}^1$, has antisymmetric $NN$ isospin components,
so it drops out the equations for the $K^- pp$ system, but remains
in the equations describing the $K^- d$ system.  All the remaining
operators form symmetric and antisymmetric pairs.
At the end there is a system of 9 (with PEST $NN$ potential) or 10
(with TSA nucleon-nucleon model) coupled operator equations, which has
the required symmetry properties.

The system of operator equations Eq. (\ref{full_oper_eq}) written in momentum space
turns into a system of integral equations. To search for a quasi-bound state
in a three-body system means to look for a solution of the homogeneous system
corresponding to Eq.~(\ref{full_oper_eq}). Calculation of three-body scattering amplitudes
require solution of the inhomogeneous system. In the both cases the integral equations
are transformed into algebraic ones. The methods of solution are different for
the quasi-bound state and scattering problems, so they are discussed in
the corresponding sections.

More details on the three-body equations with coupled $\bar{K}NN - \pi \Sigma N$
channels can be found in~\cite{myKpp_PRC} for the $K^- pp$ and in~\cite{my_Kd}
for the $K^- d$ systems.

\vspace*{2mm}
\noindent
\underline{$\bar{K}\bar{K}N - \bar{K} \pi \Sigma$ system}

As for the strangeness $S=-2$ $\bar{K}\bar{K}N$ system,
its total spin is equal to one half since an antikaon is a pseudoscalar meson. Since the
two-body interactions, namely the $\bar{K}N - \pi \Sigma$ and $\bar{K} \bar{K}$ potentials,
were chosen to have zero orbital angular momentum, the total angular momentum is also
equal to $1/2$. As in the case of the $\bar{K}NN$ system, the state of
the $\bar{K}\bar{K}N$ system with the lowest possible value of the isospin $I=1/2$
was considered.

Two identical antikaons should have a symmetric way function, therefore,
the $\bar{K}\bar{K}$ pair in $s$-wave can be in isospin one state only.
Accordingly, the three-body operators entering the AGS system for the $\bar{K}\bar{K}N$
system were symmetrized. Similarly to the $\bar{K}NN$ case, the transition operator
$X_{3,1}^1$, entering the equations describing $\bar{K}\bar{K}N$, already has the proper
symmetry properties. The remaining operators form pairs with proper symmetry properties.


\vspace*{2mm}
It is necessary to note that while Coulomb potential was directly included in
the two-body equations, used for fitting the antikaon-nucleon potentials,
the three-body calculations were performed without it (except the case of kaonic
deuterium calculations, of cause). The reason is that the expected effect of its inclusion
is small. In addition, the isospin averaged masses were used in all three-body calculations
in contrast to the two-body $\bar{K}N$ case. Accuracy of this approximation was checked
in \cite{revai_1405}, and it turned out to be quite high.

\section{Quasi-bound states}
\label{qbs_sect}

It was shown in Section \ref{V_KN_sect} that the phenomenological $\bar{K}N$ potentials
with one- and two-pole structure of the $\Lambda(1405)$ resonance and
the chirally motivated antikaon-nucleon potential can reproduce near-threshold 
experimental data on $K^- p$ scattering and kaonic hydrogen with equal accuracy.
Therefore, it is not possible to choose one of these models looking at
the two-body system only. Due to this, the three-body calculations were performed 
using all three models of the antikaon-nucleon interaction. 

The quasi-bound state in the $K^- pp$ system was the phenomenon, which
attracted the present interest to the antikaon-nucleus systems. Additionally
to being an interesting exotic object, the state could clarify
still unanswered questions on the antikaon-nucleon interaction,
in particular, the nature of the $\Lambda(1405)$ resonance.

$\bar{K}\bar{K}N$ is one more possible candidate for a strange three-body system
with the quasi-bound state in it. However, the strangeness $S=-2$ system contains
$\bar{K}\bar{K}$ interaction, which is repulsive. The question was,
whether the repulsion is strong enough to overtake $\bar{K}N$ attraction and
by this exclude the possibility of the quasi bound state formation.

\subsection{Two ways of a quasi-bound state evaluation}
\label{qbs_details_sect}

The quasi-bound state, which is a bound state with a non-zero width, for the
higher $\bar{K}NN$ (or $\bar{K}\bar{K}N$) channel, is at the same time
a resonance for the lower $\pi \Sigma N$ ($\bar{K} \pi \Sigma$) channel.
Therefore, the corresponding pole should be situated between the
$\bar{K}NN$ ($\bar{K}\bar{K}N$) and $\pi \Sigma N$ ($\bar{K} \pi \Sigma$))
thresholds on the physical energy sheet of the higher channel and on
an unphysical sheet of the lower channel. Two methods of searching of the complex
pole position by solving the homogeneous system of equations were used. 

The first one is the direct pole search with contour rotation. The correct
analytical continuation of the equations from the physical energy sheet
to the proper unphysical one is achieved by moving the momentum integration
into the complex plane. Namely, the integration was performed along a ray in
the fourth quadrant of the complex plane with some condition on the momentum
variable. After that the position $z_0$ of a quasi-bound state was found by solving
the equation ${\rm Det}(z_0) = 0$, where ${\rm Det}(z)$ is the determinant of
the linear system, obtained after discretization of the integral equations,
corresponding to Eq.~(\ref{full_oper_eq}).

Another way of a quasi-bound state searching, which avoids integration in
the complex plane, was suggested and used in~\cite{ourKNN_II}.
The idea is that every isolated and quite narrow resonance should manifest
itself at real energies. Namely, resonances are usually seen in cross-sections
of some reactions. The function $1/{\rm Det}(z)$ enters all possible amplitudes,
described by a system of three-body integral equations. Therefore, 
the function $1/|{\rm Det}(z)|^2$, entering all possible cross-sections, can be
calculated instead of some cross-sections. The function is universal, it does not
contain additional information about the particular processes in the three-body
system. The corresponding bump of the $1/|{\rm Det}(z)|^2$ function, calculated
at real energies, can be fitted by a Breit-Wigner curve with arbitrary background.
In this way an information on the resonance position and width can be obtained.

\begin{figure}
\centering
\includegraphics[width=0.8\textwidth]{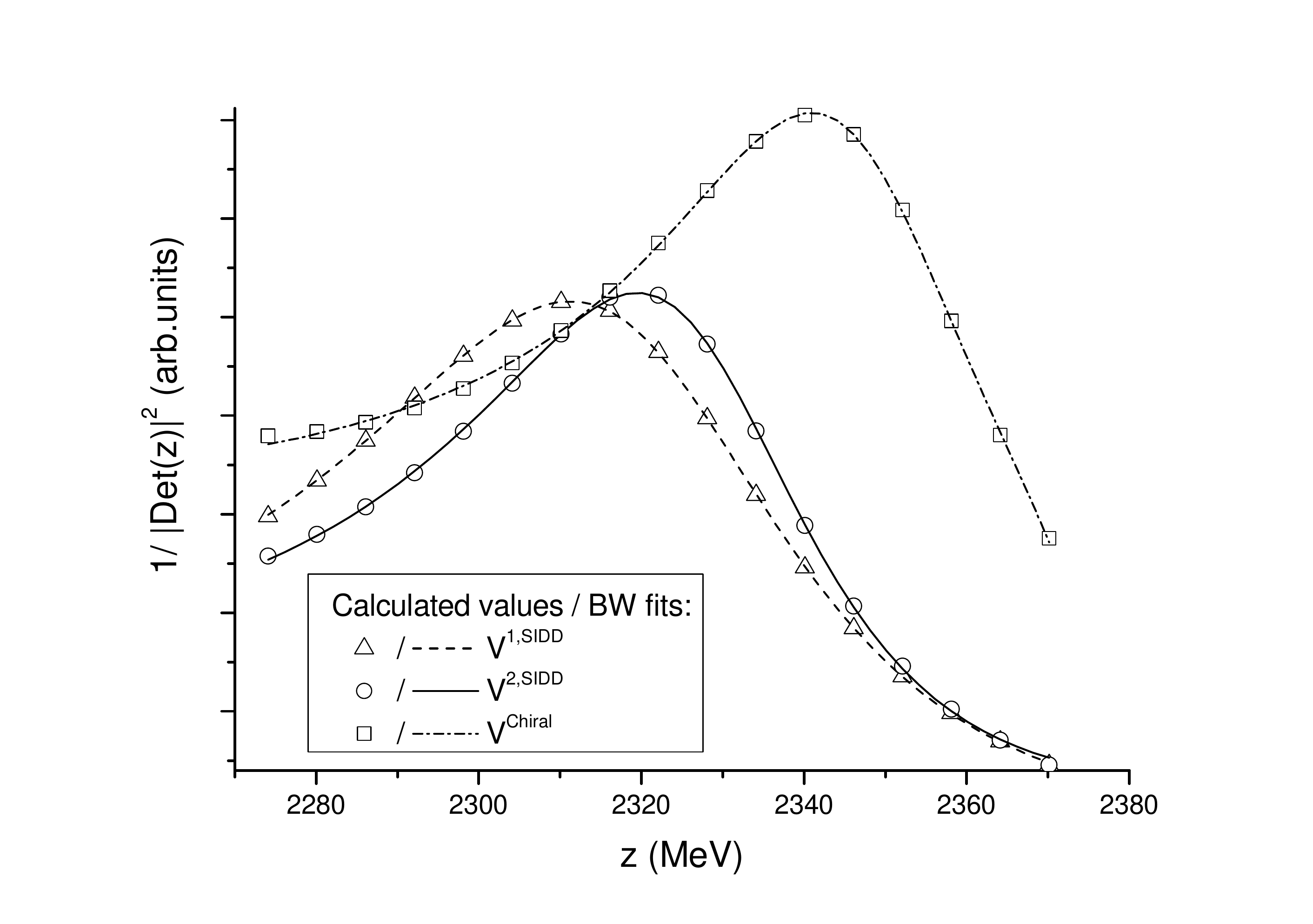}
\caption{Calculated $1/|{\rm Det}(z)|^2$ functions of the AGS system of equations
for the $K^- pp$ system (symbols) and the corresponding Breit-Wigner fits of
the obtained curves (lines) \cite{ourKNN_II} for
the one-pole $V^{\rm 1, SIDD}_{\bar{K}N - \pi \Sigma}$ (triangles and dashed line),
the two-pole $V^{\rm 2, SIDD}_{\bar{K}N - \pi \Sigma}$ (circles and solid line)
phenomenological potentials,
and the chirally motivated $V^{Chiral}_{\bar{K}N - \pi \Sigma-\pi \Lambda}$
(squares and dash-dotted line) potential.}.
\label{detsBW.fig}
\end{figure}

It is clear that the second method can work only if the resonance bump is
isolated and not too wide. The bump corresponding to the $K^- pp$ quasi-bound
state satisfies these conditions \cite{ourKNN_II}, as is seen in Fig. \ref{detsBW.fig}.
The calculated $1/|{\rm Det}(z)|^2$ functions of the AGS system of equations
are shown there as symbols while the corresponding Breit-Wigner fitting curves
are drown in lines. The results obtained with the three $\bar{K}N$ potentials,
described in Section \ref{V_KN_sect}, are shown in the figure.

Since direct search of the complex root is a non-trivial task, the Breit-Wigner values
of the $1/|{\rm Det}(z)|^2$ function can give a good starting point for it. On the
other hand, the $1/|{\rm Det}(z)|^2$ method can be used as a test of the directly
found pole position, which is free of the possible uncertainty of the proper choice
of the Riemann sheet. However, the $1/|{\rm Det}(z)|^2$ method is not easier than
the direct search, since the calculation of the determinant, which is almost equal
to the solving of a scattering problem, should be performed.

\subsection{$K^- pp$ quasi-bound state: results}
\label{qbs_results_sect}

The first dynamically exact calculation of the quasi-bound state in the $K^- pp$
system was published in~\cite{myKpp_PRL},
while the extended version of the results appeared in~\cite{myKpp_PRC}.
Existence of the $I=1/2, J^{\pi}=0^-$ three-body quasi-bound state
in the $\bar{K}NN$ system, predicted in \cite{AY1,AY2}, was confirmed there,
but the evaluated binding energy and width were strongly different.
However, the $\bar{K}N - \pi \Sigma$ potentials used in \cite{myKpp_PRL,myKpp_PRC}
do not reproduce the experimental data on the $K^- p$ system as accurately, as those
described in Section \ref{V_KN_sect}. Due to this, the calculations
devoted to the $K^- pp$ system were repeated in \cite{ourKNN_II}. The one-pole
$V^{1,SIDD}_{\bar{K}N-\pi \Sigma}$, two-pole $V^{2,SIDD}_{\bar{K}N-\pi \Sigma}$
phenomenological potentials from~\cite{my_Kd_sdvig} and the chirally motivated
$V^{Chiral}_{\bar{K}N - \pi \Sigma - \pi \Lambda}$ potential from~\cite{ourKNN_I},
described in Section \ref{V_KN_sect}, were used as an input. The other two potentials
were the two-term TSA-B NN potential \cite{Doles_NN}
together with the spin-independent exact optical $\Sigma N$ potential in isospin
$I=1/2$ state and the one-channel $V_{\Sigma N}$ in $I=3/2$,
see Section \ref{NN_SigN_sect}.

The results of the last calculations \cite{ourKNN_II} of the $K^- pp$ quasi-bound state
are shown in Table~\ref{KNN_poles.tab}. First of all, comparison of
the results obtained using the direct pole search and the $1/|{\rm Det}(z)|^2$ method
demonstrates that they are very close each to other for the every given $\bar{K}N$ potential.
Therefore, the suggested $1/|{\rm Det}(z)|^2$ method of finding mass and width of
a subthreshold resonance is efficient for the $K^- pp$ system, and the two methods
supplement each another.
\begin{center}
\begin{table}
\caption{Binding energy $B_{K^-pp}$ (MeV) and width
$\Gamma_{K^-pp}$ (MeV) of the quasi-bound state in the $K^- pp$ system \cite{ourKNN_II}:
the results of the direct pole search and of the Breit-Wigner fit of the
$1/|{\rm Det(z)}|^2$ function at real energy axis. The AGS calculations
were performed with the one-pole $V^{1,SIDD}_{\bar{K}N-\pi \Sigma}$,
two-pole $V^{2,SIDD}_{\bar{K}N-\pi \Sigma}$ phenomenological
potentials from~\cite{my_Kd_sdvig} and the chirally motivated
$V^{Chiral}_{\bar{K}N - \pi \Sigma - \pi \Lambda}$ potential
from~\cite{ourKNN_I}.}
\label{KNN_poles.tab}
\begin{center}
\begin{tabular}{ccccc}
\hline \noalign{\smallskip}
  & \multicolumn{2}{c}{Direct pole search} & \multicolumn{2}{c}{BW fit of $1/|{\rm Det(z)}|^2$} \\
\hline \noalign{\smallskip}
  &  $B_{K^- pp}$ & $\Gamma_{K^- pp}$ &  $B_{K^- pp}$ & $\Gamma_{K^- pp}$ \\
           \noalign{\smallskip} \hline \noalign{\smallskip}
 $V^{1,SIDD}_{\bar{K}N - \pi \Sigma}$ 
           & $53.3$ & $64.8$ & $54.0$ &  $66.6$ \\ \noalign{\smallskip}
 $V^{2,SIDD}_{\bar{K}N - \pi \Sigma}$
           & $47.4$ &  $49.8$ &  $46.2$ &  $51.8$ \\ \noalign{\smallskip}
 $V_{\bar{K}N - \pi \Sigma - \pi \Lambda}^{\rm Chiral}$
           & $32.2$ &  $48.6$ & $30.3$ &  $46.6$ \\
           \noalign{\smallskip} \hline
\end{tabular}
\end{center}
\end{table}
\end{center}

Another fact, seen from the results in Table~\ref{KNN_poles.tab}, is strong dependence
of the binding energy $B_{K^- pp}$ of the quasi-bound state and its width
$\Gamma_{K^- pp}$ on the $\bar{K}N$ interaction models. It was already observed in
\cite{myKpp_PRL,myKpp_PRC}, when older phenomenological antikaon-nucleon potentials
were used. In particular, it is seen from Table~\ref{KNN_poles.tab} that the quasi-bound
states resulting from the phenomenological potentials lie about $15-20$ MeV deeper than
those of the chirally motivated one. This probably is due to the energy dependence of the
chirally motivated model of the interaction. Really, all three potentials were fitted to
the experimental data near the $\bar{K}N$ threshold. When the $K^- pp$ quasi-bound state
is calculated at lower energies, the strengths of the phenomenological models of
the $\bar{K}N$ interaction are unchanged. As for the chirally motivated potential,
its energy dependence reduces the attraction at the lower energies in the $\bar{K}NN$
quasi-bound state region, thus producing the states with less binding.

The widths of the three quasi-bound states are also different: those of the two-pole models
of the $\bar{K}N$ interaction are almost coinciding, while the width evaluated using
the one-pole $V^{1,SIDD}_{\bar{K}N - \pi \Sigma}$ potential is much larger.
It is seen from Table~\ref{phys_char.tab} that the potentials with the two-pole 
$\Lambda(1405)$ structure have very close positions of the higher poles, while the pole of
the one-pole potential is different. Therefore, the difference in widths might be
connected with the different pole structure of the corresponding $\bar{K}N$ interaction
models.

Importance of the proper inclusion of the second $\pi \Sigma N$ channel in the
calculations was first demonstrated in~\cite{myKpp_PRC}. A simple complex version
of the $\bar{K}N - \pi \Sigma$ potentials, described in Section \ref{V_KN_sect},
was used in~\cite{myKpp_PRC} together with the full version with coupled channels.
This allowed to check importance of the proper inclusion of the second channel.
Comparison of the result of the one-channel complex calculation
($B^{1 \,\rm complex}_{K^- pp},\Gamma^{1 \,\rm complex}_{K^- pp}$)
with the coupled-channel one ($B^{2 \,\rm coupled}_{K^- pp},\Gamma^{2 \,\rm coupled}_{K^- pp}$)
\begin{eqnarray}
\label{2chOLDKpp}
 B^{2 \,\rm coupled}_{K^- pp} &=& 55.1 \; {\rm MeV} 
  \qquad \Gamma^{2 \,\rm coupled}_{K^- pp} = 101.8\; {\rm MeV} \\
\label{1chOLDKpp}
  B^{1 \,\rm complex}_{K^- pp} &=& 40.2 \; {\rm MeV}
  \qquad \Gamma^{1 \,\rm complex}_{K^- pp} = 77.4\; {\rm MeV}
\end{eqnarray}
shows that the quasi-bound state obtained in the full calculation with coupled channels is much
deeper and broader than the approximate one-channel one. (The values for the binding energy and
width in Eq.(\ref{2chOLDKpp}) differ from those in Table~\ref{KNN_poles.tab} since another
$\bar{K}N$ potential was used in \cite{myKpp_PRC}.) This means that the $\pi \Sigma$
channel plays an important dynamical role in forming the three-body quasi-bound state, over
its obvious role of absorbing flux from the $\bar{K}N$ channel. Thus, proper inclusion of
the second $\pi \Sigma$ channel is crucial for the $\bar{K}NN$ system.

It was found later in~\cite{ourKNN_II} that use of the exact optical
$\bar{K}N$ potential can serve an alternate way of direct inclusion
of the $\pi \Sigma$ channel. An accuracy of use of the exact
optical $\bar{K}N$ potential, which gives exactly the same on- and off-shell elastic
$\bar{K}N$ amplitude as the original potential with coupled channels, was checked
in one-channel AGS calculations for the three actual $\bar{K}N$ potentials.
The ``exact optical'' binding energies differ only slightly from the full coupled-channel results
from  Table~\ref{KNN_poles.tab}, while the widths gain more visible error:
\begin{eqnarray}
 B_{K^- pp}^{1,SIDD,Opt} &=& 54.2\; {\rm MeV} 
  \qquad  \Gamma_{K^- pp}^{1,SIDD,Opt} = 61.0 \; {\rm MeV} \\
 B_{K^- pp}^{2,SIDD,Opt} &=& 47.4\; {\rm MeV} 
  \qquad  \Gamma_{K^- pp}^{2,SIDD,Opt} = 46.0 \; {\rm MeV}  \\
 B_{K^- pp}^{Chiral,Opt} &=& 32.9; {\rm MeV}
  \qquad  \Gamma_{K^- pp}^{Chiral,Opt} = 48.8 \; {\rm MeV.}
\end{eqnarray}
However, the difference in widths is not dramatic, so the one-channel
Faddeev calculation with the exact optical $\bar{K}N$ potential could be quite satisfactory
approximation to the full calculation with coupled channels.

\subsection{$K^- pp$ quasi-bound state: comparison to other results}
\label{qbs_comparison_sect}

The three binding energy $B_{K^- pp}$ and width $\Gamma_{K^- pp}$ values of 
the $K^- pp$ quasi-bound state, shown in Table \ref{KNN_poles.tab}, can be compared
with worth mentioning other theoretical results. Those are: the original prediction
of the deep and narrow quasi-bound $K^- pp$ state~\cite{AY1,AY2}, the results obtained
in the earlier Faddeev calculation~\cite{myKpp_PRC}, the most recent results of alternative
calculation using the same equations with different input~\cite{ikedasato} and
two variational results~\cite{vaizeyap,evrei}. Only calculations presented
in~\cite{ikedasato} together with the earlier ones~\cite{myKpp_PRL,myKpp_PRC}
were performed with directly included $\pi \Sigma N$ channel. All others
take it into account approximately. The second problem is that none of the $\bar{K}N$
potentials, used in all other  $K^- pp$ calculations, reproduce data on near-threshold
$K^- p$ scattering with the same level of accuracy as those described in Section~\ref{V_KN_sect}.
In addition, none of them reproduce the $1s$ level shift of kaonic hydrogen directly.

The binding energy of the quasi-bound $K^- pp$ state and its width were obtained in~\cite{AY1,AY2}
from a G-matrix calculation, which is a many-body technics. The one-channel simple
complex $\bar{K}N$ potential used in those calculations does not reproduce the actual
$K^- p$ experimental data. Finally, the authors of~\cite{AY1,AY2} take into
account only the $\bar{K}NN$ channel. In a funny way all the defects of the calculation
presented in~\cite{AY1,AY2} led to the binding energy ($48$ MeV) and width ($61$ MeV),
which are quite close to the exact results from Table~\ref{KNN_poles.tab}
obtained with the two-pole and the one-pole phenomenological potentials, respectively.

The earlier result for the binding energy $B_{K^- pp} = 55.1$ MeV~\cite{myKpp_PRC}
is very close to the actual one from Table~\ref{KNN_poles.tab} calculated with the one-pole
phenomenological $\bar{K}N - \pi \Sigma$ potential. In fact, in both cases the same
three-body equations with coupled $\bar{K}NN$ and $\pi \Sigma N$ channels
were solved. In addition, the same model of the antikaon-nucleon interaction was used,
but with different sets of parameters. This difference influenced the width of
the quasi-bound state: the older $\Gamma_{K^- pp} = 100.2$ MeV is much larger.

Coupled-channel AGS equations were also solved in~\cite{ikedasato}
with chirally motivated energy dependent and independent $\bar{K}N$
potentials. Therefore, in principle, those results obtained with the energy
dependent version of the $\bar{K}N$ potential $V^{E-dep}_{\bar{K}N}$
should give a result, which is close to those from Table~\ref{KNN_poles.tab}
with chirally motivated model of interaction
$V_{\bar{K}N - \pi \Sigma - \pi \Lambda}^{\rm Chiral}$. However, only
those width ($34-46$ MeV) is comparable to the $\Gamma_{K^- pp}^{\rm Chiral}$,
while the binding energy obtained in~\cite{ikedasato} ($9-16$ MeV) is much smaller
than the actual one from Table~\ref{KNN_poles.tab}. The situation is opposite
when the actual results are compared with those obtained
in \cite{ikedasato} using the energy independent antikaon-nucleon potential. Namely,
binding energies reported in that paper ($44-58$ MeV) are comparable with 
those from Table~\ref{KNN_poles.tab}  evaluated using phenomenological
$V^{1,SIDD}_{\bar{K}N}$ and $V^{2,SIDD}_{\bar{K}N}$
potentials, however, their widths $34-40$ MeV are much smaller.

The authors of \cite{ikedasato} neglect the $\Sigma N$ interaction in their
calculations. It was shown in \cite{myKpp_PRL,myKpp_PRC} that dependence of
the three-body $K^- pp$ pole position on $\Sigma N$ is weak. However, 
when the interaction is switched off completely, like in the case of \cite{ikedasato},
some visible effect manifests itself.

An approximation used in the chirally motivated models used in \cite{ikedasato}
is more serious reason of the difference. Namely, the energy-dependent square root
factors, responsible for the correct normalization of the $\bar{K}N$ amplitudes,
are replaced by constant masses. This can be reasonable for the highest $\bar{K}N$ channel,
however, it is certainly a poor approximation for the lower $\pi \Sigma$ and
$\pi \Lambda$ channels. The role of this approximation in the AGS calculations
was checked in \cite{ourKNN_II}, the obtained approximate binding energy
$25$ MeV is really much smaller than the original one $32$ MeV, presented in
Table~\ref{KNN_poles.tab}. The remaining difference between the results
from Table \ref{KNN_poles.tab} and those from \cite{ikedasato} could be explained
by the lower accuracy of reproducing experimental $K^- p$ data by the $\bar{K}N$
potentials from \cite{ikedasato}.

Finally, no second pole in the $K^- pp$ system reported in~\cite{ikedasato} was found
in \cite{ourKNN_II}. The search was performed with all three $\bar{K}N$ potentials
in the corresponding energy region (binding energy $67-89$ MeV and width $244-320$ MeV).

Variational calculations, performed by two groups, reported the results which
are very close to those obtained in~\cite{ikedasato} with the energy-dependent
potential:
$B_{K^- pp} = 17-23$ MeV, $\Gamma_{K^- pp} = 40-70$ MeV in~\cite{vaizeyap}
and 
$B_{K^- pp} = 15.7$ MeV, $\Gamma_{K^- pp} = 41.2$ MeV in~\cite{evrei}.
However, there are a few problematic points in~\cite{vaizeyap,evrei}.
First of all, the variational calculations were performed solely in the
$\bar{K}NN$ channel. The authors of the variational calculations
used a one-channel $\bar{K}N$ potential, derived from a chirally motivated model
of interaction with many couped channels. However, the potential is not
the ``exact optical'' one. In fact, it is not clear, how this one-channel potential
is connected to the original one and whether it still reproduces some experimental
$K^- p$ data.

Moreover, the position of the $K^- pp$ quasi-bound state was determined
in~\cite{vaizeyap,evrei} using only the real part of this complex $\bar{K}N$ potential,
as a real bound state. The width was estimated as the expectation value of
the imaginary part of the potential. This, essentially perturbative, treatment of
the inelasticity might be justified for quite narrow resonances, but the
$K^- pp$ quasi-bound state is certainly not of this type.

Another serious problem of the variational calculations is their method of
treatment of the energy dependence of the $\bar{K}N$ potential in the
few-body calculations. It was already shown in the previous subsection
that the energy dependence of the chirally motivated model of the $\bar{K} N$ 
interaction is very important for the $K^- pp$ quasi-bound state position.
While momentum space Faddeev integral equations allow the exact
treatment of this energy dependence, variational calculations in coordinate space
can use only energy independent interactions. Due to this the energy of the $\bar{K}N$
potential was fixed in ~\cite{vaizeyap,evrei} at a ``self-consistent'' value $z_{\bar{K}N}$.

A series of calculations using the exact AGS equations was performed in \cite{ourKNN_II}
with differently fixed two-particle $\bar{K}N$ energies $z_{\bar{K}N}$ in the couplings
of the chirally motivated interaction. The conclusion was, that it is not possible
to define an ``averaged'' $z_{\bar{K}N}$, for which the fixed-energy
chirally motivated interaction, even in the correct three-body calculation, can
yield a correct $K^- pp$ quasi-bound state position. 
First, the calculations of \cite{ourKNN_II} show, that a real $z_{\bar{K}N}$ has
absolutely no chance to reproduce or reasonably approximate the exact quasi-bound
state position, even with correct treatment of the imaginary part of the interaction, unlike
in~\cite{vaizeyap,evrei}. Second, the way, how the ``self-consistent''
value of (generally complex) $z_{\bar{K}N}$ is defined in the papers does not seem
to guarantee, that the correct value will be reached or at least approximated.
In view of the above considerations, the results of~\cite{vaizeyap,evrei}
can be considered as rough estimates of what a really energy dependent $\bar{K}N$
interaction will produce in the $K^- pp$ system.

After publication of the exact results~\cite{ourKNN_II} one more paper on
the $K^-pp$ system appeared \cite{gruziny}. Hyperspherical harmonics in
the momentum representation and Faddeev equations in configuration space
were used there. However, the authors collected all defects of other
approximate calculations. In particular, they used simple complex antikaon-nucleon
potentials and, therefore, neglected proper inclusion of the $\pi \Sigma N$
channel, which is crucial for the system. In addition, the $\bar{K}N$ potentials are
those from \cite{AY2, vaizeyap}, which have problems with reproducing of
the experimental $K^- p$ data. Keeping all this in mind, the results of \cite{gruziny}
hardly can be reliable.

\subsection{$K^- d$ quasi-bound state}
\label{qbsKd_results_sect}

The strongly attractive isospin-zero part of the $\bar{K}N$ potential plays less
important role in the spin-one $K^- d$ state of the $\bar{K}NN$ system than
in the spin-zero $K^- pp$. Therefore, if a quasi-bound state exists in $K^-d$,
it should have smaller binding energy than in $K^- pp$. The Faddeev calculations
of the $K^- d$ scattering  length $a_{K^- d}$, described in Section~\ref{Kd_elastic_sect},
gave some evidence that such a state could exists~\cite{ourKNN_I}.
A simple analytical continuation of the effective range formula below the $K^- d$
threshold suggests a $K^- d$ quasi-bound state with binding energy $14.6 - 19.6$ MeV
(the energy is measured from the $K^- d$ threshold) and width $15.6 - 22.0$ MeV
for the three antikaon-nucleon potentials from Section~\ref{V_KN_sect}.

However, a systematic search for these states, performed in \cite{ourKNN_I} with the
same two-body input as for the $K^- pp$ system, did not find the corresponding poles
in the complex energy plane between the $\pi \Sigma N$ and $K^- d$ thresholds.
The reason of discrepancy between the effective range estimations and the direct calculations
must be the validity of the effective range formula, which is limited to the vicinity
of the corresponding threshold. Since the $K^- d$ state is expected to have, similarly
to $K^- pp$, rather large width, it is definitely out of this region.

It was demonstrated in~\cite{ourKNN_I} that increasing of the attraction in
isospin-zero $\bar{K}N$ subsystem by hands (in the phenomenological antikaon-nucleon
potentials only) leads to appearing of $K^- d$ quasi-bound states. Therefore,
the isospin-zero attraction in the $\bar{K}N$ system is not strong enough to
bind antikaon to deuteron. It is necessary to note that the $K^- d$ system with
strong two-body interactions only is considered here. An atomic state caused by Coulomb
interaction, kaonic deuterium, exists and will be considered later.

\subsection{$\bar{K}\bar{K}N$ system: results}
\label{KKN_res_sect}

The calculations of the quasi-bound state in the $\bar{K}\bar{K}N$ system
were performed with the two $\bar{K}\bar{K}$ interactions 
described in Section~\ref{V_KK_sect} (Original $V_{\bar{K}\bar{K}}^{Orig}$ and
Lattice-motivated $V_{\bar{K}\bar{K}}^{Latt}$) 
and three $\bar{K}N$ potentials from Section~\ref{V_KN_sect}: the phenomenological
one-pole $V^{1,SIDD}_{\bar{K}N - \pi \Sigma}$ and
two-pole $V^{2,SIDD}_{\bar{K}N - \pi \Sigma}$ phenomenological potentials together
with the chirally-motivated potential $V^{Chiral}_{\bar{K}N - \pi \Sigma - \pi \Lambda}$.
The results are presented in Table \ref{KKN_poles.tab}.
It turned out that all combinations of the two-body interactions
lead to a quasi-bound state in the three-body $\bar{K}\bar{K}N$ system.
The quasi-bound state exists in the strangeness $S=-2$ system
in spite of the repulsive character of the $\bar{K}\bar{K}$ interaction
Comparison with the $K^- pp$ characteristics from Table~\ref{KNN_poles.tab}
shows that the quasi-bound state in the strangeness $S=-2$
$\bar{K}\bar{K}N$ system is much shallower and broader than the one in the
$S=-1$ $K^- pp$ system for the given $\bar{K}N$ potential.
\begin{center}
\begin{table}
\caption{Binding energy $B_{\bar{K}\bar{K}N}$ (MeV) and width
$\Gamma_{\bar{K}\bar{K}N}$ (MeV) of the quasi-bound state in the $\bar{K}\bar{K}N$
system \cite{my_KKN}: the results of the direct pole search and of the Breit-Wigner
fit of the $1/|{\rm Det(z)}|^2$ function at real energy axis. The AGS calculations
were performed with the one-pole $V^{1,SIDD}_{\bar{K}N-\pi \Sigma}$,
two-pole $V^{2,SIDD}_{\bar{K}N-\pi \Sigma}$ phenomenological
potentials from~\cite{my_Kd_sdvig} and the chirally motivated
$V^{Chiral}_{\bar{K}N - \pi \Sigma - \pi \Lambda}$ potential
from~\cite{ourKNN_I}. $V^{Orig}_{\bar{K}\bar{K}}$ and
$V^{Lattice}_{\bar{K}\bar{K}}$ models of the antikaon-antikaon interaction were used.}
\label{KKN_poles.tab}
\begin{center}
\begin{tabular}{cccccc}
\hline \noalign{\smallskip}
  & & \multicolumn{2}{c}{Direct pole search} & \multicolumn{2}{c}{BW fit of $1/|{\rm Det(z)}|^2$} \\
\hline \noalign{\smallskip}
  & &  $B_{\bar{K}\bar{K}N}$ & $\Gamma_{\bar{K}\bar{K}N}$ 
    &  $B_{\bar{K}\bar{K}N}$ & $\Gamma_{\bar{K}\bar{K}N}$ \\
           \noalign{\smallskip} \hline \noalign{\smallskip}
 & $V^{1,SIDD}_{\bar{K}N - \pi \Sigma}$
           & $11.9$ &  $102.2$ & $17.1$ &  $110.8$ \\ \noalign{\smallskip}
$V^{Orig}_{\bar{K}\bar{K}}$ and & 
      $V^{2,SIDD}_{\bar{K}N - \pi \Sigma}$
           & $23.1$ &  $91.4$ &  $23.7$ &  $77.6$ \\ \noalign{\smallskip}
 &  $V_{\bar{K}N - \pi \Sigma - \pi \Lambda}^{\rm Chiral}$
           & $15.5$ &  $63.5$ & $15.9$ &  $57.4$ \\
           \noalign{\smallskip} \hline \noalign{\smallskip}
 & $V^{1,SIDD}_{\bar{K}N - \pi \Sigma}$
           & $19.5$ &  $102.0$ & $23.7$ &  $103.7$ \\ \noalign{\smallskip}
$V^{Lattice}_{\bar{K}\bar{K}}$ and & 
      $V^{2,SIDD}_{\bar{K}N - \pi \Sigma}$
           & $25.9$ &  $84.6$ &  $26.4$ &  $76.8$ \\ \noalign{\smallskip}
 &  $V_{\bar{K}N - \pi \Sigma - \pi \Lambda}^{\rm Chiral}$
           & $16.1$ &  $61.3$ & $15.9$ &  $60.0$ \\
           \noalign{\smallskip} \hline
\end{tabular}
\end{center}
\end{table}
\end{center}

Two methods of the quasi-bound state evaluation were used: the direct
search method and the Breit-Wigner fit of the inverse determinant.
It is seen from the Table \ref{KKN_poles.tab} that the accuracy of the inverse
determinant method is much lower for the phenomenological $\bar{K}N$ interactions
than for the chirally motivated one (and for the $K^- pp$ system too). The reason
is the larger widths of the ''phenomenological'' $\bar{K}\bar{K}N$ states, which
means that the corresponding bumps are less pronounced, so they hardly can be
fitted reliably by Breit-Wigner curves.

The found $\bar{K}\bar{K}N$ quasi-bound state has the same quantum numbers as
a $\Xi$ baryon with $J^P=(1/2)^+$. The available experimental information on
the $\Xi$ spectrum is rather limited, see PDG \cite{PDG}. There is a $\Xi$(1950)
listed by the PDG, but its quantum numbers $J^P$ are not determined, and it is 
unclear whether it should be identified with the quark-model state. It is possible
that there are more than one resonance in this region. However,
in spite of the fact, that the $\Xi(1950)$ state is situated above
the $\bar{K}\bar{K}N$ threshold, four of the experimental values would be roughly
consistent with the quasi-bound state found in the calculation~\cite{my_KKN}.
Specifically, the experiment reported in Ref.~\cite{Dauber:1969} yielded a mass
$1894 \pm 18$ MeV and a width $98 \pm 23$ MeV that is compatible with the range
of values for the evaluated pole position.

An investigation on the $\bar{K}\bar{K}N$ system was also performed
in \cite{KanadaEn'yo:2008wm}, but several uncontrolled approximations
were done there. In particular, energy-independent as well as 
energy-dependent potentials were used, but the two-body energy of the latter was fixed
arbitrarily. Moreover, the imaginary parts of all complex potentials
were completely ignored in the variational calculations in \cite{KanadaEn'yo:2008wm},
the widths of the state were estimated separately. As a result, the binding energies
are compared to the exact ones from \cite{my_KKN}, but the widths of
the $\bar{K}\bar{K}N$ state are strongly underestimated.

\section{Near-threshold $K^- d$ scattering}
\label{Kd_elastic_sect}

\subsection{Methods and exact results}
\label{Kd_scatt_exact_sect}

The $K^- pp$ quasi-bound state is a very interesting exotic object. However,
it is not clear whether the accuracy of experimental results will be enough
to draw some conclusions from comparison of the data with theoretical predictions.
No strong quasi-bound state was found in the $\bar{K}NN$ system with other quantum
numbers $K^- d$ \cite{ourKNN_I}, but an atomic state, kaonic deuterium, exists,
and its energy levels can be accurately measured. In addition, scattering of an antikaon
on a deuteron can be studied.

Exact calculations of the near-threshold $K^- d$ scattering were performed 
in \cite{my_Kd_sdvig,ourKNN_I} using the three antikaon-nucleon potentials,
described in Section \ref{V_KN_sect}, and different versions of the $\Sigma N$ and $NN$
potentials, described in Section~\ref{NN_SigN_sect}. Namely, the exact optical and
a simple complex versions of the  spin-dependent and spin-independent
$\Sigma N -\Lambda N$ potentials were used. The calculations were performed with TSA-A,
TSA-B, and PEST models of the $NN$ interaction.

The inhomogeneous system of the integral AGS equations, corresponding to
Eq.(\ref{full_oper_eq}) and describing the $K^- d$ scattering, was transformed into
the system of algebraic equations. It is known, that the original, one-channel, integral
Faddeev equations have moving logarithmic singularities in the kernels
when scattering above a three-body breakup threshold is considered. The $K^- d$
amplitudes were calculated from zero up to the three-body breakup $\bar{K}NN$ threshold,
so, in principle, the equations could be free of the singularities. However, the lower
$\pi \Sigma N$ channel is opened when the $K^- d$ scattering is considered, which
causes appearance of the logarithmic singularities even below the three-body 
breakup $\bar{K}NN$ threshold. The problem was solved by interpolating of the unknown solutions
in the singular region by certain polynomials and subsequent analytical integrating
of the singular part of the kernels.

The $K^- d$ scattering lengths $a_{K^- d}$ obtained with the
one- $V^{1,SIDD}_{\bar{K}N-\pi \Sigma}$ and two-pole
$V^{2,SIDD}_{\bar{K}N-\pi \Sigma}$ versions of the phenomenological $\bar{K}N$ potential
in \cite{my_Kd_sdvig} together with the chirally-motivated potential
$V^{Chiral}_{\bar{K}N - \pi \Sigma - \pi \Lambda}$ in \cite{ourKNN_I}
are shown in Table~\ref{Kd_aReff_EGam.tab}. It is seen, that the chirally motivated
potential leads to slightly larger absolute value of the real and the imaginary
part of the scattering length than the phenomenological ones. However, the difference
is small, so the three different models of the $\bar{K}N$ interaction, which reproduce
the low-energy data on the $K^- p$ scattering and kaonic hydrogen with the same level
of accuracy, give quite similar results for low-energy $K^- d$ scattering.
It means that it is not possible to solve the question of the number
of the poles forming the $\Lambda(1405)$ resonance from the results
on the near-threshold elastic $K^- d$ scattering.

The small difference between the ``phenomenological'' and ``chiral'' results
of the $a_{K^- d}$ calculations
is opposite to the results obtained for the $K^- pp$ system (see Section~\ref{qbs_results_sect}),
where the binding energy and width of the $K^- pp$ quasi-bound state
were calculated using the same equations (the homogeneous ones with properly changed
quantum numbers, of cause) and input. In that case the three-body observables
obtained with the three $\bar{K}N$ potentials turned out to be very different each
from the other. The reason of this difference between the results for the near-threshold
scattering and the quasi-bound state calculations could be the fact, that while
the $a_{K^- d}$ values were calculated near the $\bar{K}NN$ threshold, the $K^- pp$ pole
positions are situated far below it.

The amplitudes of the elastic $K^- d$ scattering for kinetic energy from $0$ to $E_{\rm deu}$,
calculated using the three versions of the $\bar{K}N$ potential, are shown in Fig.3
of \cite{my_Kd_sdvig} and in Fig 5 of \cite{ourKNN_I} in a form of $k \cot \delta(k)$ function.
The chosen representation demonstrates that the elastic near-threshold $K^- d$
amplitudes can be approximated by the effective range expansion rather accurately
since the lines are almost straight. The calculated effective ranges $r^{\rm eff}_{K^- d}$ 
of the $K^- d$ scattering, evaluated using the obtained $K^- d$ amplitudes, are shown
in Table~\ref{Kd_aReff_EGam.tab}.

The dependence of the full coupled-channel results on the $NN$ and $\Sigma N (-\Lambda N)$
interaction models was investigated in \cite{my_Kd}. The antikaon-nucleon
phenomenological potentials, used there, reproduce the earlier KEK data on kaonic hydrogen
and not the actual ones by SIDDHARTA. However, the results, obtained
with those phenomenological $\bar{K}N$ potentials, are relative, so they must be valid
for the actual potentials as well. In order to investigate dependence of the three-body
results on the $NN$ model of interaction, TSA-A, TSA-B, and PEST nucleon-nucleon
potentials were used. It turned out that the difference for the $K^- d$ scattering length
is very small even for the potentials with and without repulsion at short distances
(TSA and PEST, respectively). Therefore, the $s$-wave $NN$ interaction
plays a minor role in the calculations. Most likely, it is caused by the relative weakness
of the $NN$ interaction as compared to the $\bar{K}N$ one. Indeed, the quasi bound state
in the latter system (which is the $\Lambda(1405)$ resonance with $E_{\bar{K}N} \approx -23$ MeV)
is much deeper than the deuteron bound state ($E_{\rm deu} \approx 2$ MeV). Due to this,
some visible effect from higher partial waves in $NN$ is also not expected.

The dependence of $a_{K^- d}$ on the $\Sigma N (-\Lambda N)$ interaction was also investigated
in \cite{my_Kd}. The $K^- d$ scattering lengths were calculated with the exact optical
and the simple complex versions of the spin dependent $V^{\rm Sdep}$ and spin independent
$V^{\rm Sind}$ potentials. The results obtained with the two versions of the $\Sigma N (-\Lambda N)$
potential  $V^{\rm Sdep}$ and $V^{\rm Sind}$ in exact optical
form are very close, while their simple complex versions are slightly different.
However, the largest error does not exceed $3 \%$, therefore, the dependence of
the $K^- d$ scattering length $a_{K^- p}$ on the $\Sigma N - (\Lambda N)$ interaction is also weak.

\subsection{Approximate calculations and comparison to other results}
\label{Kd_approx_compare_sect}

It is hard to make a comparison with other theoretical results due to different
methods and inputs used there. Due to this, several approximate calculations,
in particular, one-channel $\bar{K}NN$ calculations with a complex
and the exact optical $\bar{K}N$ potentials, were performed in \cite{my_Kd}.
In addition, a so-called FCA method was tested there. 

In order to investigate the importance of the direct inclusion of the $\pi \Sigma N$
channel, the one-channel AGS calculations were performed in \cite{my_Kd}. It means that
Eq.~(\ref{full_oper_eq}) with $\alpha = \beta = 1$ were solved, therefore,
only the $\bar{K}N$ and $NN$ $T$-matrices enter the equations. The exact optical
and two simple complex one-channel $\bar{K}N (- \pi \Sigma)$ potentials approximating
the full coupled-channel one- and two-pole phenomenological models of the interaction were used.
As written in Section \ref{V_KN_sect}, the exact optical potential $V^{\rm{Opt}}$ provides
exactly the same elastic $\bar{K}N$ amplitude as the coupled-channel
model of the interaction. Its energy-dependent strength parameters are defined by
Eq.~(\ref{lambdaOpt}) with $\bar{\alpha},\bar{\beta} = 1,2$ stands for the $\bar{K}N$
and $\pi \Sigma$ channels, respectively.

The complex constants of the simple complex potentials were obtained in two ways.
The first version of the simple complex $\bar{K}N$ potential reproduces the $K^- p$
scattering length $a_{K^- p}$ and the pole position $z_{1}$ of the corresponding
coupled-channel version of the potentials. The second one gives the same isospin
$I_i=0$ and $I_i=1$ $\bar{K}N$ scattering lengths as the full $\bar{K}N - \pi \Sigma$ potential.

It was found in \cite{my_Kd} that the one-channel AGS calculation with
the exact optical $\bar{K}N$ potential, giving exactly the same elastic $\bar{K}N$
amplitude as the corresponding coupled-channel phenomenological potential,
is the best approximation. Its error does not exceed $2$ percents. 
(The same is true for the results obtained with the chirally motivated $\bar{K}N$ potential,
see \cite{ourKNN_I}.)
On the contrary, the both simple complex $\bar{K}N$ potentials led to very inaccurate
three-body results. Therefore, the one-channel Faddeev-type calculation with a simple
complex antikaon-nucleon potential is not a good approximation for the low-energy
elastic $K^- d$ scattering. 

One more approximate method, used for the $a_{K^- d}$ calculations, is a
so-called ``Fixed center approximation to Faddeev equations'' (FCA), introduced
in~\cite{Kd_KOR}. In fact, it is a variant of FSA or a two-center formula.
The fixed-scatterer approximation (FSA) or a two-center problem assumes,
that the scattering of a projectile particle takes place on two much heavier target
particles, separated by a fixed distance. The motion of the heavy particles is
subsequently taken into account by averaging of the obtained projectile-target amplitude
over the bound state wave function of the target. The approximation is well known and
works properly in atomic physics, where an electron is really
much lighter than a nucleon or an ion. Since the antikaon mass is
just a half of the mass of a nucleon, it was expected, that FSA hardly can be a
good approximation for the $K^- d$ scattering length calculation.

The derivations of the FCA formula from Faddeev equations presented in~\cite{Kd_KOR}
already rises questions, while the proper derivations of the FSA formula was done
in~\cite{peresypkin}. The accuracy of the FCA was checked in \cite{my_Kd} using the same
input as in the AGS equations in order to make the comparison as adequate as possible.

First of all, the $\bar{K}N$ scattering lengths provided by the coupled-channel
$\bar{K}N - \pi \Sigma$ potentials together with the deuteron wave function, corresponding
to the TSA-B $NN$ potential, were used in the FCA formula. Second, all $\bar{K}^0 n$
parts were removed from the formula because they drop off the AGS system of equations
after the antisymmetrization. Finally, the fact, that the FCA formula was obtained for a local
$\bar{K}N$ potential, while the separable $\bar{K}N - \pi \Sigma$ potentials were used in
the Faddeev equations, was took into account, and the corresponding changes in the FCA
formula were made.

The results of using of the FCA formula without ''isospin-breaking effects'' stay far away
from the full calculation. While the errors for the imaginary part are not so large,
the absolute value of the real part is underestimated by about $30 \%$. Therefore,
the calculations performed in \cite{my_Kd} show that FCA is a poor approximation
for the $K^- d$ scattering length calculation. It is also seen from the figure that
the accuracy is lower for the two-pole model of the $\bar{K}N$ interaction.

Therefore, among the approximate results the FCA was demonstrated to be
the least accurate approximation, especially in reproducing of the real part
of the $K^- d$ scattering length. On the contrary, the one-channel AGS calculation
with the exact optical $\bar{K}N (-\pi \Sigma)$ potential gives the best approximation
to the full coupled-channel result. All approximations are less accurate for
the two-pole phenomenological model of the $\bar{K}N - \pi \Sigma$ interaction.

Calculations of the $K^- d$ scattering length were performed by other authors
using Faddeev equations in~\cite{Kd_BFMS_new,Kd_TGE,Kd_TDD,Kd_Deloff}, while
the FCA method was used in~\cite{Kd_KOR,ruzecky}.
The result of the very recent calculation with coupled channels~\cite{Kd_BFMS_new}
has real part of $a_{K^- d}$, which almost coincides with the result for chirally
motivated potential shown in Table  \ref{Kd_aReff_EGam.tab}.
The imaginary part of the $K^- d$ scattering length from~\cite{Kd_BFMS_new} is
slightly larger. It might be caused by the fact that the model of the $\bar{K}N$
interaction, used there, was not fitted to the kaonic hydrogen data directly, but
through the $K^- p$ scattering length and the Deser-type approximate formula, which 
has larger error for the imaginary part of the level shift.

The two old $a_{K^- d}$ values~\cite{Kd_TDD,Kd_TGE}, obtained within
coupled-channel Faddeev approach, significantly underestimate the imaginary
part of the $K^- d$ scattering length, while their real parts are rather close to those
in Table~\ref{Kd_aReff_EGam.tab}.

One more result of a Faddeev calculation~\cite{Kd_Deloff} lies far away
from all the others with very small absolute value of the real part of $a_{K^- d}$.
One of the reasons is that the $K^- d$ scattering length
was obtained in~\cite{Kd_Deloff} from one-channel Faddeev equations with
a complex $\bar{K}N$ potential. However, the underestimation of the absolute value
of its real part in comparison to other Faddeev calculations is so large, that it
cannot be explained by the method only. The additional reason of the difference
must be the $\bar{K}N$ potential, used in the paper. It gives so high position of
the $K^- p$ quasi bound state ($1439$ MeV), that it is situated above the $K^- p$ threshold.

The $a_{K^- d}$ values of~\cite{Kd_KOR} obtained using FCA method differ significantly
from all other results. The absolute value of the real part of $a_{K^- d}$ from~\cite{Kd_KOR}
and its imaginary part are too large, which is caused by two factors. The first one is the
FCA formula itself, which was shown to be inaccurate for the present system. The second reason
are too large $\bar{K}N$ scattering lengths, used as the inputs.

The result of \cite{ruzecky} was obtained by simple applying of two approximate
formulas: FCA and the corrected Deser formula, used for calculation of the $\bar{K}N$
scattering lengths, entering the FCA. The values of~\cite{ruzecky} suffer not only from
the cumulative errors from the two approximations, but from using of the DEAR results
on kaonic hydrogen $1s$ level shift and width as  well. Indeed, it was already written
that the error of the corrected Deser formula makes about $10 \%$ for two-body case,
the accuracy of the FCA was shown to be poor. As for the problems with DEAR experimental
data, they were demonstrated in \cite{our_KN} and in other theoretical works.

\section{$1s$ level shift of kaonic deuterium}
\label{kaonic_deu_sect}

The shift of the $1s$ level in the kaonic deuterium (which, strictly speaking, is the
``antikaonic'' deuterium) and its width are caused by the presence of the strong
interactions in addition to the Coulomb one. It is a directly measurable value, which
is free of a few uncertainties connected with an experiment on the $K^- pp$ quasi-bound
state. However, from theoretical point of view it is harder task due to necessity to take
Coulomb potential into account directly together with the strong ones.

There are two ways to solve three-body problems accurately:
solution of Faddeev equations or use of variational methods.
However, for the case of an hadronic atom both methods face serious difficulties.
The problem of the long range Coulomb force exists in the Faddeev
approach, while variational methods suffer from the presence of two very
different distance scales, which both are relevant for the calculations.

Due to this, at the first step the $1s$ level energy of the kaonic deuterium was
calculated approximately using a two-body model of the atom.
At the next step a method for simultaneous treatment of a short range plus Coulomb forces
in three-body problems based on Faddeev equations~\cite{Papp1} was used,
and the lowest level of kaonic deuterium was calculated dynamically exactly.

\subsection{Approximate calculation of kaonic deuterium $1s$ level}
\label{Kdshift_approx_sect}

The approximate calculation of the kaonic deuterium was performed assuming
that the atom can be considered as a two-body system consisting of a point-like deuteron,
interacting with an antikaon through a complex strong $K^- - d$ potential and Coulomb.
By this the size of a deuteron was taken into account only effectively through
the strong potential, which reproduces the elastic three-body $K^- d$ amplitudes,
evaluated before. Keeping in mind the relative values of a deuteron and Bohr radius
of the kaonic deuterium, the approximation seemed well grounded.

The complex two-body $K^- - d$ potential, constructed and used for investigation
of the kaonic deuterium by Lippmann-Schwinger equation, is a two-term 
separable potential
\begin{equation}
\label{VKd}
V_{K^- d}(\vec{k},\vec{k}') = \lambda_{1,K^- d} \, g_1(\vec{k}) g_1(\vec{k}')
+ \lambda_{2,K^- d} \, g_2(\vec{k}) g_2(\vec{k}')
\end{equation}
with Yamaguchi form-factors
\begin{equation}
\label{gKd}
g_i(k) = \frac{1}{\beta_{i,K^- d}^2 + k^2}, \qquad i=1,2.
\end{equation}
The complex strength parameters $\lambda_{1,K^- d}$ and $\lambda_{2,K^- d}$
were fixed by the conditions, that the $V_{K^- d}$ potential reproduces
the $K^- d$ scattering length $a_{K^- d}$ and the effective range
$r^{\rm eff}_{K^- d}$, obtained with one of the $\bar{K}N - \pi \Sigma$
potentials and presented in Table \ref{Kd_aReff_EGam.tab}. A variation of the real
$\beta_{1,K^- d}$ and $\beta_{2,K^- d}$ parameters allowed to reproduce the full
near-threshold $K^- d$ amplitudes from \cite{my_Kd_sdvig,ourKNN_I} more accurately.
As a result, the near-threshold amplitudes obtained from the three-body
calculations $f^{(3)}_{K^- d}$ are reproduced by the two-body
$K^- - d$ potentials through the interval $[0,E_{\rm deu}]$ with such
accuracy, that the two-body functions $k \cot \delta^{(2)}(k)$
are indistinguishable from the three-body $k \cot \delta^{(3)}(k)$.

The parameters of the potentials are shown in Table 3 of~\cite{my_Kd_sdvig}
and in Eqs.(18,19) of~\cite{ourKNN_I}. Both $\beta_{1,K^- d}$ and
$\beta_{2,K^- d}$ parameters for every $K^- - d$ potential are much
smaller than the corresponding $\beta^{\bar{K}N}$ parameter of the $\bar{K}N$
potential.
The constructed two-body complex potentials $V_{K^- d}$ were used
in the Lippmann-Schwinger equation. The calculations of the binding energy of
a two-body system, described by the Hamiltonian with the strong and
Coulomb interactions were performed in the same way as those of the $K^- p$
system, see Section~\ref{V_KN_sect}.
\begin{center}
\begin{table}
\caption{Scattering length $a_{K^- d}$ (fm) and effective range
$r^{eff}_{K^- d}$ (fm) of $K^- d$ system obtained from AGS calculations
with the one-pole $V^{1,SIDD}_{\bar{K}N-\pi \Sigma}$, two-pole 
$V^{2,SIDD}_{\bar{K}N-\pi \Sigma}$ phenomenological potentials and
the chirally-motivated $V^{Chiral}_{\bar{K}N - \pi \Sigma - \pi \Lambda}$
potential. Approximate results for the $1s$ level shift $\Delta E^{K^- d}_{1s}$
(eV) and width $\Gamma^{K^- d}_{1s}$ (eV) of kaonic deuterium, corresponding
to the AGS results on the near-threshold elastic amplitudes, are also shown.}
\label{Kd_aReff_EGam.tab}
\begin{center}
\begin{tabular}{ccccc}
\hline \noalign{\smallskip}
$\bar{K}N$ potential used & $a_{K^- d}$ & $r^{eff}_{K^- d}$ 
  & $\Delta E^{K^- d}_{1s}$ & \quad $\Gamma^{K^- d}_{1s}$\\
  \hline \noalign{\smallskip}
 $V^{1,SIDD}_{\bar{K}N}$  & $-1.49 + i \, 1.24$ 
   \qquad & $0.69 -  i \, 1.31$ & -785 & 1018 \\ \noalign{\smallskip}
 $V^{2,SIDD}_{\bar{K}N}$  & $-1.51 + i \, 1.25$ 
   \qquad & $0.69 -  i \, 1.34$ & -797 & 1025 \\ \noalign{\smallskip}
 $V_{\bar{K}N}^{\rm Chiral}$ & $-1.59 +  i \, 1.32$ 
  \qquad & $0.50 -  i \, 1.17$ & -828 & 1055 \\ 
\noalign{\smallskip} \hline
\end{tabular}
\end{center}
\end{table}
\end{center}

The shifts $\Delta E^{K^- d}_{1s}$ and widths $\Gamma^{K^- d}_{1s}$
of the $1s$ level of kaonic deuterium, corresponding to the three models
of the $\bar{K}N$ interaction, described in Section~\ref{V_KN_sect}, are shown
in Table \ref{Kd_aReff_EGam.tab}. It is seen that the ``chirally motivated''
absolute values of the level shift $\Delta E_{1s}^{K^- d}$ and the width
$\Gamma_{1s}^{K^- d}$ are both slightly larger than those obtained using
the phenomenological $\bar{K}N - \pi \Sigma$ potentials.
However, all three results do not differ one from the other more than several
percents. It is similar to the case of the $K^- d$ scattering length calculations,
which turned out to be very close for the three $\bar{K}N$ potentials.
The important point here is the fact that all three $\bar{K}N$ potentials
reproduce the low-energy experimental data on $K^- p$ scattering and kaonic hydrogen
with the same level of accuracy. It is also important that the $1s$ level
of kaonic deuterium is situated not far from the $\bar{K}NN$ threshold.

The closeness of the results for kaonic deuterium means that comparison
of the theoretical predictions with eventual experimental results hardly could
choose one of the models
of the $\bar{K}N$ interaction, especially taking into account the large widths
$\Delta E^{K^- d}_{1s}$. Therefore, it could not be possible to say, whether
the potential of the antikaon-nucleon interaction should have one- or two-pole
structure of the $\Lambda(1405)$ resonance and whether the potential should
be energy dependent or not. It is seen from Tables~\ref{phys_char.tab}
and \ref{Kd_aReff_EGam.tab} that there is no correlation between the pole
or poles of the $\Lambda(1405)$ resonance given by a $\bar{K}N $
potential and the three-body $K^- d$ elastic scattering or kaonic deuterium
characteristics obtained using the potential.

Inaccuracy of the corrected Deser formula Eq.(\ref{corDes}) was already shown
for the two-body $K^- p$ system, but some authors use it for the kaonic deuterium
as well. Due to this, an accuracy of the formula was checked for this three-body system.
The results were obtained using the $a_{K^- d}$ values from Table~\ref{Kd_aReff_EGam.tab}.
Being compared to the $\Delta E_{K^- d}$ and $\Gamma_{K^- d}$ from the same
table, the "corrected Deser" results show large error for all three versions
of the antikaon-nucleon interaction. While difference for the shift is not so
drastic, the width of the $1s$ level of the kaonic deuterium is underestimated
by the corrected Deser formula by $\sim 30\%$.

The $1s$ level shift and width presented in Table \ref{Kd_aReff_EGam.tab}
are not exact, they were evaluated using the two-body approximation,
which, however, is well-grounded. Information on the three-body strong part
is taken into account indirectly through the $K^- - d$
potential, reproducing the exact elastic $K^- d$ amplitudes. On the contrary,
the corrected Deser formula contains no three-body information at all since
the only input is a $K^- d$ scattering length, which is a complex number. Moreover,
the formula relies on further approximations, which are absent in the accurate
approximate calculations.

\subsection{Exact calculation of kaonic deuterium: 
Faddeev equations with Coulomb interaction}
\label{papp_eq_sect}

Exact calculations of the kaonic deuterium were performed using
a method~\cite{Papp1} for simultaneous treatment of short range plus Coulomb forces
in three-body problems, based on Faddeev equations. 
The method was successfully applied for purely Coulomb systems with
attraction and repulsion and for the short range plus repulsive Coulomb
forces. The case of an hadronic atom with three strongly interacting particles and
Coulomb attraction between certain pairs was not considered before. 

The basic idea of the method is to transform the Faddeev integral equations into a matrix
form using a special discrete and complete set of Coulomb Sturmian functions
as a basis. Written in coordinate space the Coulomb Sturmian functions are orthogonal with
the weight function $1/r$. So that they form a bi-orthogonal and complete set
with their counter-parts. The most remarkable feature of this particular set is, that
in this representation the matrix of the two-body $(z-h^c)$ operator, where $z$ is
an energy and $h^c$ is the pure two-body Coulomb Hamiltonian, is tridiagonal.
When  this property is used for evaluation of the matrix elements of the two-body
Coulomb Green's function $g^c$, an infinite tridiagonal set of equations, which can be
solved exactly, is obtained. The same holds for the matrix elements of the free two-body
Green's function $g^0$.

The system of equations with the Coulomb and strong interactions was solved in \cite{our_Kdexact}
for kaonic deuterium. This calculation is different from all other three-body calculations,
described before. Already the initial form of the Faddeev equations for the kaonic
deuterium differs from those for the pure strong interactions,
described in Section~\ref{AGS_sect}. First, the equations should be written
in coordinate space, while the AGS equations were written in momentum
space. Second, since the Coulomb interaction acts between $K^-$ and the proton,
the particle basis was used and not the isospin one. Finally, the Faddeev equations with
Coulomb do not define the transition operators, as e.g. those in Eq.(\ref{U_coupled}),
but the wave functions.

The equations are written in the Noble form~\cite{Noble}, when the Coulomb interaction
appears in the Green's functions. As usual for Faddeev-type equations, there are three
partition channels $\alpha = (pn,K^-)$,  $(pK^-,n)$, $(nK^-,p)$ and three sets of Jacobi
coordinates. The system of homogeneous equations to be solved contains the matrix elements of
the overlap between the basis functions from different Jacobi coordinate sets
and of the strong potentials. They all can be calculated directly. The remaining
parts of the kernel are matrix elements of the three-body partition Green's
functions $G_{\alpha}$. They are the basic quantities of the method, and their
calculation depends on the partition channel.

The partition Green function $G_{(pK^-,n)}$ of the $(pK^-,n)$ channel
contains Coulomb interaction in its ``natural'' coordinate. It describes
the $(pK^-)$ subsystem and the neutron, which
do not interact between themselves. Due to this, $G_{(pK^-,n)}$ can be
calculated taking a convolution integral along a suitable contour in the complex
energy plane over two two-body Green functions. As for the matrix elements
of the two-body Green functions, they can be calculated using the properties
of the Coulomb Sturmian basis and solving a resolvent equation.

The situation with the remaining $G_{\alpha}$ functions is more complicated.
In the case of the $\alpha = (pn,K^-)$ and $(nK^-,p)$ channels
the Coulomb interaction is written not in its ``natural'' coordinates.
Due to this, it should be rewritten as a sum of the Coulomb potential
in the natural coordinates plus a short range potential $U_{\alpha}$,
which is a ``polarization potential''. The three-body Green function
$G_{\alpha}^{ch}$ containing Coulomb potential in natural for the channel
coordinates is called the ''channel Green function'', and it is evaluated similarly
to the previous $\alpha = (pK^-,n)$ case. Namely, since the function describes
a two-body subsystem and the non-interacting with it third particle,
the $G_{(pn,K^-)}^{ch}$ and $G_{(nK^-,p)}^{ch}$ functions can be found
by taking a convolution integral with two two-body Green's functions.
At the last step the $G_{\alpha}$ function is found from the equation,
containing the obtained channel Green function $G_{\alpha}^{ch}$ and
the polarization potential $U_{\alpha}$
\begin{equation}
\label{Geq}
 G_{\alpha}(z) = G^{ch}_{\alpha}(z) + G^{ch}_{\alpha}(z) U_{\alpha}
 G_{\alpha}(z).
\end{equation}

For the kaonic deuterium calculations it was necessary to take the isospin
dependence of the $\bar{K}N$ interaction into account. In particle representation
it means that the strong $V_{pK^-}^s$ potential is a $2 \times 2$ matrix,
containing $V^s_{pK^-,pK^-}$, $V^s_{pK^-,n\bar{K}^0}$ and $V^s_{n\bar{K}^0,n\bar{K}^0}$
elements. Due to this, the final equations for the kaonic deuterium have
four Faddeev components, including the additional one in the $(n\bar{K}^0,n)$
channel.

The solution of the Faddeev-type equations with Coulomb gave the full energy of
the $1s$ level. Since the aim of \cite{our_Kdexact} was evaluation of the $1s$ level
shift of kaonic deuterium caused by the strong interactions between the antikaon
and the nucleons, it was necessary to define the energy, from which the real part
of the shift is measured. It can be the lowest eigenvalue of the channel Green function
or of the ``original'' Green function of the $(pn,K^-)$ channel. The first one corresponds
to a deuteron and an antikaon feeling a Coulomb force from the center of mass
of the deuteron. The second reflects the fact that the antikaon interacts via Coulomb
force not with the center of the deuteron, but with the proton. In principle,
the correct one should be the second variant, however, all approximate
approaches use an analogy of the first one as the basic point, due to this
it was used in \cite{our_Kdexact} as well. In any case, the difference between
both versions is small.

\subsection{Exact calculation of kaonic deuterium: results}
\label{Kd_accurate_sect}

The calculation performed in~\cite{our_Kdexact} was considered as
a first test of the method for the description of three-body hadronic atoms.
Due to this, the second three-body particle channel $\pi \Sigma N$ was not
directly included and no energy dependent potentials (exact optical
or chirally motivated one) were used. The $\bar{K}N$ and $NN$ interactions were
described using one-term separable complex potentials
with Yamaguchi form factors. Four versions of the $\bar{K}N$ potential
$V_I, V_{II}, V_{III}$ and $V_{IV}$, used in the calculations, give the $1s$ level
shift of the kaonic hydrogen within or close to the SIDDHARTA data and a reasonable
fit to the  elastic $K^- p \to K^- p$ and charge exchange $K^- p \to \bar{K}^0 n$ cross-sections.
Parameters of the potentials can be found in Table I of~\cite{our_Kdexact}.
The nucleon-nucleon potential reproduces the $NN$ scattering lengths, low-energy phase shifts
and the deuteron binding energy in the $np$ state.
\begin{table*}
\caption{Exact $1s$ level shifts $\Delta E$ (eV) and widths $\Gamma$ (eV)
of the kaonic deuterium for the four complex $\bar{K}N$ potentials $V_I, V_{II}, V_{III}$,
and $V_{IV}$. The approximate results obtained using the corrected Deser formula and
the complex $K^- - d$ potential are also shown.}
\label{Kd_shiftExact.tab}
\begin{center}
\begin{tabular}{ccccccc}
\hline \noalign{\smallskip}
  & \multicolumn{2}{c}{Corrected Deser} & \multicolumn{2}{c}{Complex $V_{K^- - d}$} &  
   \multicolumn{2}{c}{Exact Faddeev} \\
\noalign{\smallskip} \hline \noalign{\smallskip}
 & $\Delta_{1s}^{K^- d}$ & $\Gamma_{1s}^{K^- d}$ & 
   $\Delta_{1s}^{K^- d}$ & $\Gamma_{1s}^{K^- d}$ & 
   $\Delta_{1s}^{K^- d}$ & $\Gamma_{1s}^{K^- d}$ \\
\noalign{\smallskip} \hline \noalign{\smallskip}
$V_I$     &  $-675$ & $702$  & $-650$ & $868$   & $-641$ & $856$\\
$V_{II}$  &  $-694$ & $740$  & $-658$ & $920$   & $-646$ & $888$\\
$V_{III}$ &  $-795$ & $780$  & $-747$ & $1034$  & $-732$ & $980$\\
$V_{IV}$  &  $-750$ & $620$  & $-740$ & $844$   & $-736$ & $826$\\
\noalign{\smallskip} \hline
\end{tabular}
\end{center}
\end{table*}

The results of the dynamically exact calculations of the kaonic deuterium are presented
in Table~\ref{Kd_shiftExact.tab}. The absolute values of the $1s$ level shift were
found in the region $641 - 736$ eV, while the width variates between $826 - 980$ eV.
Both observables are smaller than the accurate results from~\cite{ourKNN_I}, shown in
Table~\ref{Kd_aReff_EGam.tab}. However, it is necessary to remember that both
calculations differ not only by the three-body methods, but also by the two-body input.
To make the comparison reasonable, the two-body approximate calculation,
described in Section~\ref{Kdshift_approx_sect}, was repeated with the simple
complex $\bar{K}N$ potentials $V_I, V_{II}, V_{III}$, and $V_{IV}$. The corrected Deser
formula was also checked for these potentials. The approximate results are shown
in Table~\ref{Kd_shiftExact.tab}.

It is seen that the two-body approximate calculation, described in
Section~\ref{Kdshift_approx_sect}, makes $\le 2 \%$ error for the shift and $\le 5 \%$ 
for the width, so it is quite accurate. It is an expected result keeping in mind the relative
values of deuteron and Bohr radius of kaonic deuterium. The corrected Deser formula
Eq.(\ref{corDes}) leads to $2 - 8 \%$ error in the shift, and strongly, up to $25 \%$,
underestimates the width.

It is also possible to compare the approximate results obtained in \cite{our_Kdexact}
with the four complex $\bar{K}N$ potentials $V_I, V_{II}, V_{III}$, and $V_{IV}$
and in \cite{ourKNN_I} with the coupled-channel models of the antikaon-nucleon interaction
(the phenomenological $V^{1,SIDD}_{\bar{K}N-\pi \Sigma}$ and $V^{2,SIDD}_{\bar{K}N-\pi \Sigma}$
with one- and two-pole structure of the $\Lambda(1405)$ resonance respectively, and
the chirally motivated $V^{Chiral}_{\bar{K}N - \pi \Sigma-\pi \Lambda}$).
It is seen that the `'complex one-channel'' absolute values of the $1s$ level shift
and width shown in Table~\ref{Kd_shiftExact.tab} are smaller than the `'coupled-channel''
ones presented in Table \ref{Kd_aReff_EGam.tab}. The similar situation was observed
with the exactly evaluated characteristics of the strong pole in the $K^- pp$ system,
while a one-channel simple complex antikaon - nucleon potential led to more
narrow and less bound quasi-bound state than the coupled-channel version
(see Eqs.(\ref{2chOLDKpp},\ref{1chOLDKpp})). But the differences for the kaonic deuterium
are smaller than those for the $K^- pp$ quasi-bound state.

The very recent calculations~\cite{revai_exactKd} of the kaonic deuterium
$1s$ level shift were performed using the same Faddeev-type equations with Coulomb
interaction as in \cite{our_Kdexact}, but with energy-dependent $\bar{K}N$ potentials. 
Namely, the exact optical versions of the one- and two-pole phenomenological
$\bar{K}N - \pi \Sigma$  potentials and of the chirally motivated $\bar{K}N-\pi \Sigma - \pi \Lambda$
interaction model were used. The predicted $1s$ level shifts ($800 \pm 30$ eV) and widths
($960 \pm 40$ eV) are larger by absolute value than the exact ones from Table~\ref{Kd_shiftExact.tab}
evaluated with the simple complex antikaon nucleon potentials.

Keeping in mind good accuracy of the results obtained with the exact optical
$\bar{K}N$ potentials for all three-body $\bar{K}NN$ observables, demonstrated in the present
paper, the predictions of \cite{revai_exactKd} for the kaonic deuterium must be the most
accurate ones up to date. The two-body approximation used in \cite{my_Kd_sdvig,ourKNN_I},
being compared to the more accurate approach of \cite{revai_exactKd}, gives very accurate
value of the $1s$ level shift (the error is $\le 2 \%$), while the error for the width is
larger ($\le 9 \%$).

\section{Summary}
\label{Summary_sect}

The three-body antikaon nucleon systems could provide an important information about
the antikaon nucleon interaction. It is quite useful since the two-body $\bar{K}N$ potentials
of different type can reproduce all low-energy experimental data with the same level
of accuracy. This fact was demonstrated on the example of the phenomenological $\bar{K}N-\pi \Sigma$
potentials with one and two-pole structure of the $\Lambda(1405)$ resonance together with
the chirally motivated $\bar{K}N - \pi \Sigma - \pi \Lambda$ potential. Being used
in the three-body calculations, the three $\bar{K}N$ potentials allowed to investigate
the influence of the $\bar{K}N$ model on the results.

It was found that while the quasi-bound state position in the $K^- pp$ and $K^- K^- p$ systems
strongly depends on the model of the $\bar{K}N$ interaction, the near-threshold observables
($K^-d$ scattering length, elastic near-threshold $K^- d$ amplitudes, $1s$ level shift and
width of kaonic deuterium) are almost insensitive to it. Therefore, some conclusions on
the number of poles of the $\Lambda(1405)$ resonance could be done only if a hight accuracy
measurement of $K^- pp$ binding energy and width will be done. Probably, one of the existing
experiments: by HADES~\cite{HADES} and LEPS~\cite{LEPS} collaborations, and in J-PARC
\cite{J-PARC_E15,J-PARC_E27} hopefully will clarify the situation with the $K^- pp$ quasi-bound
state, - will do it. While dependence of the three-body results on the $\bar{K}N$ potentials
is different for the different systems and processes, dependence on $NN$ and $\Sigma N$ interactions
is weak in all cases.

Comparison of the exact results with some approximate ones revealed the most accurate
approximations. In particular, the one-channel Faddeev calculations give results, which are 
very close to the coupled-channel calculations if the exact optical $\bar{K}N$ potential
is used. This fact gives a hope for four-body calculations, which are already very complicated
without additional coupled-channel structure. It is necessary to note here that the ''exact optical''
potential is defined as an energy dependent potential, which exactly reproduces the elastic amplitudes
of the corresponding potential with coupled channels.

As for the kaonic deuterium, influenced mainly by Coulomb interaction, the shift of its $1s$ level
caused by the strong interactions is described quite accurately in the two-body approximation.
The $K^- - d$ complex potential should herewith reproduce the exact elastic three-body $K^- d$
amplitudes, and the Lippmann-Schwinger equation must be solved exactly with Coulomb plus the strong
potentials. Of cause, the exact calculation is more precise and, therefore, is preferable.
The predicted by the exact calculations $1s$ level energy could be checked by SIDDHARTA-2
collaboration \cite{SIDDHARTA2}.

The suggested $1/|{\rm Det(z)}|^2$ method of theoretical evaluation of an underthreshold
resonance is quite accurate for rather narrow and well pronounced resonances. It could supplement
the direct search of the pole providing the first estimation and working as a control. The method
is free from the uncertainties connected with the calculations on the complex plane, but it has
the logarithmic singularities in the kernels of the integral equations.

The next step in the field of the few-body systems consisting of antikaons and nucleons
should be done toward the four body systems. They could give more possibilities, but theoretical
investigations of them are much more complicated.


\end{document}